\documentclass[onecolumn]{article}%
\usepackage{amsmath}
\usepackage{amsfonts}
\usepackage{amssymb}
\usepackage{graphicx}
\usepackage{revsymb}%
\newtheorem{theorem}{Theorem}

\newenvironment{proof}[1][Proof]{\noindent\textbf{#1.} }{\ \rule{0.5em}{0.5em}}
\begin{document}

\title{Reality Conditions for Spin Foams}
\author{Suresh. K.\ Maran}
\maketitle

\begin{abstract}
An idea of reality conditions in the context of spin foams (Barrett-Crane
models) is developed. The square of areas are the most elementary observables
in the case of spin foams. This observation implies that simplest reality
conditions in the context of the Barrett-Crane models is that the all possible
scalar products of the bivectors associated to the triangles of a four simplex
be real. The continuum generalization of this is the area metric reality
constraint: the area metric is real iff a non-degenerate metric is real or
imaginary. Classical real general relativity (all signatures) can be extracted
from complex general relativity by imposing the area metric reality
constraint. The Plebanski theory can be modified by adding a Lagrange
multiplier to impose the area metric reality condition to derive classical
real general relativity. I discuss the $SO(4,C)$ BF model and $SO(4,C)$
Barrett-Crane model. It appears that the spin foam models in 4D for all the
signatures are the projections of the $SO(4,C)$ spin foam model using the
reality constraints on the bivectors.

\end{abstract}
\tableofcontents

\section{Introduction}

The Ashtekar self-dual formalism \cite{AABOOK} in its most intuitive form
relates to complex general relativity. A set of conditions referred to as
reality conditions \cite{AABOOK} is to be imposed to go to real general
relativity. The reality conditions not only impose the reality of the physics,
but also the Lorentzian signature \cite{AABOOK}. The goal of this article is
to introduce analogous idea of reality conditions for the covariant version of
quantum gravity - the Barrett-Crane model \cite{BCReimmanion}.

The Barrett-Crane model \cite{BCReimmanion} is defined by a set of constraints
defined on the bivectors associated to the triangles of a four simplex. The
constraints need to be realized at the quantum level to define a quantum state
of a tetrahedron or to assign an amplitude to a four simplex. Let us make the
bivectors to be complex which corresponds to $SO(4,C)$ general relativity.
Then we need a set of conditions to extract Barrett-Crane models for real
general relativity. In the Barrett-Crane models the elementary physical
observables are the square of area eigenvalues associated to the triangles. So
the simplest reality conditions in Barrett-Crane models is to be defined in
terms of the squares of the areas. This is equivalent to expecting the inner
product of a bivector with itself to be real. For real general relativity we
also need the square of areas corresponding to the sum of the bivectors of the
triangles to be real. As will be explained in this article this is equivalent
to expecting the inner product of the bivectors corresponding to any pair of
triangles to be real.

The continuum generalization of the squares of the areas of triangles of
simplicial manifold is the Area metric. It can be shown that the reality of an
area metric is equivalent to the reality of a geometry. An area metric can be
defined as an inner product of a bivector two-form field with itself. Because
of this the reality of an area metric can be imposed using a Lagrange
multiplier in the Plebanski formulation of $SO(4,C)$ general relativity. The
Barrett-Crane model corresponds to the discrete analog of the Plebanski
formalism. Analogously, it can be shown that the reality of the bivector inner
products discussed before is the discrete equivalent of the area metric
reality constraint.

An attempt by me to rigorously develop and unify the various models for the
Lorentzian general relativity was made in Ref:\cite{MyRigorousSpinFoam}. The
attempt was made to derive the two models by directly solving the
Barrett-Crane constraints. The Barrett-Crane cross-simplicity constraint
operator was explicitly written using the Gelfand-Naimarck representation
theory of $SL(2,C)$ \cite{IMG}. But after numerous attempts I\ could not
obtain any solution for the constraint. But the efforts in this research lead
to the development of the reality conditions for the spin foam models. Then as
will be discussed in this article the Barrett-Crane models for real general
relativity theories for all signatures appears to be related to that of the
$SO(4,C)$ general relativity through the quantum version of the discretized
area metric reality condition\footnote{The Barrett-Crane model based on the
propagators on the null-cone \cite{BCLorentzian} is an exception to this. This
needs to be carefully investigated.}. In this way we have a unified
understanding of the Barrett-Crane models for the four dimensional real
general relativity theories for all signatures (non-degenerate) and the
$SO(4,C)$ general relativity.

The layout of this article is as follows:

\begin{itemize}
\item Section two: I\ review the Plebanski formulation \cite{Plebanski} of
$SO(4,C)$ general relativity starting from vectorial actions.

\item Section three: I discuss the area metric reality constraint. After
solving the Plebanski (simplicity) constraints\cite{Plebanski}, I\ show that,
the area metric reality constraint requires the space-time metric to be real
or imaginary for the non-denegerate case. I\ modify the vectorial Plebanski
actions by adding a Lagrange multiplier to impose the reality constraint.

\item Section four: I discuss the discretization of the area metric reality
constraint on the simplicial manifolds in the context of the Barrett-Crane
theory \cite{BCReimmanion}.

\item Section five: I discuss the spin foam model for the $SO(4,C)$ BF theory.

\item Section six: I discuss the $SO(4,C)$ Barrett-Crane model.

\item Section seven: Using the bivector scalar product reality constraint the
Barrett-Crane models for the real general relativity for all signatures and
$SO(4,C)$ general relativity are discussed in a unified manner.

\item Appendices: I\ discuss the necessary representation theories. I\ discuss
the area metric reality constraint for arbitrary metrics. I also discuss the
field theory over group formalism for the $SO(4,C)$ general relativity.
\end{itemize}

\section{ $SO(4,C)$ General Relativity}

Plebanski's work \cite{Plebanski} on complex general relativity presents a way
of recasting general relativity in terms of bivector 2-form fields instead of
tetrad fields \cite{APAL} or space-time metrics. It helped to reformulate
general relativity as a topological field theory called the BF theory with a
constraint (for example Reisenberger \cite{MPR1}). Originally Plebanski's work
was formulated using spinors instead of vectors. The vector version of the
work can be used to formulate spin foam models of general relativity
\cite{MPR1}, \cite{ClassActFoam}. Understanding the physics behind this theory
simplifies with the use of spinors. Here I\ would like to review the Plebanski
theory for a $SO(4,C)$ general relativity on a four dimensional real manifold
starting from vectorial actions.

Let me define some notations to be used in this article. I would like to use
the letters $i,j,k,l,m,n$ as $SO(4,C)$ vector indices, the letters
$a,b,c,d,e,f,g,h$ as space-time coordinate indices, the letters $A,B,C,D,E,F$
as spinorial indices to do spinorial expansion on the coordinate indices.

In the cases of Riemannian and $SO(4,C)$ general relativity the Lie algebra
elements are the same as the bivectors. On arbitrary bivectors $a^{ij}$ and
$b^{ij}$, I define $\footnote{The wedge product in the bivector coordinates
plays a critical role in the spin foam models. This is the reason why the
$\wedge$ is used to denote a bivector product instead of an exterior
product.}$
\begin{align*}
a\wedge b  &  =\frac{1}{2}\epsilon_{ijkl}a^{ij}b^{kl}~~\text{and}\\
a\bullet b  &  =\frac{1}{2}\eta_{ik}\eta_{jl}a^{ij}b^{kl}.
\end{align*}

Consider a four dimensional manifold $M$. Let $A$ be a $SO(4,C)$ connection
1-form and $B^{ij}$ a complex bivector valued $2$-form on M. I\ would like to
restrict myself to non-denegerate general relativity in this and the next
section by assuming $b=\frac{1}{4!}\epsilon^{abcd}B_{ab}\wedge B_{cd}\neq0$.
Let $F$ be the curvature 2-form of the connection $A$. I define real and
complex continuum \ $SO(4,C)$ BF\ theory actions as follows,%
\begin{equation}
S_{cBF}(A,B_{ij})=\int_{M}\varepsilon^{abcd}B_{ab}\wedge F_{cd}~~\text{and}
\label{BFcmplx}%
\end{equation}%
\begin{equation}
S_{rBF}(A,B_{ij},\bar{A},\bar{B}_{ij})=\operatorname{Re}\int_{M}%
\varepsilon^{abcd}B_{ab}\wedge F_{cd}. \label{BFreal}%
\end{equation}
The $S_{cBF}$ is considered as a holomorphic functional of it's variables. In
$S_{rBF}$ the variables $A,B_{ij}$ and their complex conjugates are considered
as independent variables. The wedge is defined in the Lie algebra coordinates.
The field equations corresponding to the extrema of these actions are
\begin{align*}
D_{[a}B_{bc]}  &  =0\text{~and}\\
F_{cd}  &  =0.
\end{align*}

$BF$ theories are topological field theories. It is easy to show that the
local variations of solutions of the field equations are gauged out under the
symmetries of the actions \cite{JCBintro}.

The Plebanski actions for $SO(4,C)$ general relativity is got by adding a
constraint term to the BF actions. First let me define a complex action
\cite{MPR1},%
\begin{equation}
S_{cGR}(A,B_{ij},\phi)=\int_{M}\left[  \varepsilon^{abcd}B_{ab}\wedge
F_{cd}+\frac{1}{2}b\phi^{abcd}B_{ab}\wedge B_{cd}\right]  d^{4}x,
\label{GRactionComplex}%
\end{equation}
and a real action
\begin{equation}
S_{rGR}(A,B_{ij},\phi,\bar{A},\bar{B}_{ij},\bar{\phi})=\operatorname{Re}%
S_{C}(A,B_{ij},\phi). \label{ComplexAction}%
\end{equation}
The complex action is a holomorphic functional of it's variables. Here $\phi$
is a complex tensor with the symmetries of the Riemann curvature tensor such
that $\phi^{abcd}\epsilon_{abcd}=0$. The $b$ is inserted to ensure the
invariance of the actions under coordinate change.

The field equations corresponding to the extrema of the actions $S_{C}$ and
$S$ are
\begin{subequations}
\label{field equation}%
\begin{align}
D_{[a}B_{bc]}^{ij}  &  =0\text{,}\label{Beq}\\
\frac{1}{2}\varepsilon^{abcd}F_{cd}^{ij}  &  =b\phi^{abcd}B_{cd}^{ij}\text{
and,}\label{Feq}\\
B_{ab}\wedge B_{cd}-b\epsilon_{abcd}  &  =0\text{,} \label{CSeq}%
\end{align}
where $D$ is the covariant derivative defined by the connection $A$. The field
equations for both the actions are the same.

Let me first discuss the content of equation (\ref{CSeq}) called the
simplicity constraint. The $B_{ab}$ can be expressed in spinorial form as
\end{subequations}
\[
B_{ab}^{ij}=B_{AB}^{ij}\epsilon_{\acute{A}\acute{B}}+B_{\acute{A}\acute{B}%
}^{ij}\epsilon_{AB},
\]
where the spinor $B_{AB}$ and $B_{\acute{A}\acute{B}}$ are considered as
independent variables. The tensor
\[
P_{abcd}=B_{ab}\wedge B_{cd}-b\epsilon_{abcd}%
\]
has the symmetries of the Riemann curvature tensor and it's pseudoscalar
component is zero. In appendix A the general ideas related to the spinorial
decomposition of a tensor with the symmetries of the Riemann Curvature tensor
have been summarized. The spinorial decomposition of $P_{abcd}$ is given by%
\begin{align*}
P_{abcd}  &  =B_{(AB}\wedge B_{CD)}\epsilon_{\acute{A}\acute{B}}%
\epsilon_{\acute{C}\acute{D}}+B_{(\acute{A}\acute{B}}\wedge B_{\acute{C}%
\acute{D})}\epsilon_{AB}\epsilon_{CD}+\\
&  \frac{\tilde{b}}{6}\frac{\delta_{c[a}\delta_{b]d}}{2}+B_{AB}\wedge
B_{\acute{A}\acute{B}}(\epsilon_{\acute{A}\acute{B}}\epsilon_{CD}%
+\epsilon_{AB}\epsilon_{\acute{C}\acute{D}}),
\end{align*}
where $\tilde{b}=B_{AB}\wedge B^{AB}+B_{\acute{A}\acute{B}}\wedge B^{\acute
{A}\acute{B}}$. Therefore the spinorial equivalents of the equations
(\ref{CSeq}) are
\begin{subequations}
\label{eq.Pleb}%
\begin{align}
B_{(AB}\wedge B_{CD)}  &  =0,\\
B_{(\acute{A}\acute{B}}\wedge B_{\acute{C}\acute{D})}  &  =0,\\
B_{AB}\wedge B^{AB}+B_{\acute{A}\acute{B}}\wedge B^{\acute{A}\acute{B}}  &
=0~~\text{and}\\
B_{AB}\wedge B_{\acute{A}\acute{B}}  &  =0.
\end{align}
These equations have been analyzed by Plebanski \cite{Plebanski}. The only
difference between my work (also Reisenberger \cite{MPR1}) and Plebanski's
work is that I\ have spinorially decomposed on the coordinate indices of $B$
instead of the vector indices. But this does not prevent me from adapting
Plebanski's analysis of these equations as the algebra is the same. From
Plebanski's work, we can conclude that the above equations imply $B_{ab}%
^{ij}=$ $\theta_{a}^{[i}\theta_{b}^{j]}$ where $\theta_{a}^{i}$ are a complex
tetrad. Equations (\ref{eq.Pleb}) are not modified by changing the signs of
$B_{AB}$ or/and $B_{\acute{A}\acute{B}}$. These are equivalent to replacing
$B_{ab}$ by $-B_{ab}$ or $\pm\frac{1}{2}\epsilon_{ab}^{cd}B_{cd}$ which
produce three more solution of the equations \cite{LFKK}, \cite{MPR1}.

The four solutions and their physical nature were discussed in the context of
Riemannian general relativity by Reisenberger \cite{MPR1}. It can be shown
that equation (\ref{Beq}) is equivalent to the zero torsion$\ $
condition\footnote{For a proof please see footnote-7 in Ref.\cite{MPR1}.}.
Then $A$ must be the complex Levi-Civita connection of the complex metric
$g_{ab}=\delta_{ij}\theta_{a}^{i}\theta_{b}^{j}$ on $M$. Because of this the
curvature tensor $F_{ab}^{cd}=F_{ab}^{ij}\theta_{i}^{c}\theta_{j}^{c}$
satisfies the Bianchi identities. This makes $F$ to be the $SO(4,C)$ Riemann
Curvature tensor. Using the metric $g_{ab}$ and it's inverse $g^{ab}$ we can
lower and raise coordinate indices.

Let me assume I have solved the simplicity constraint, and $dB=0$. Substitute
in the action $S$ the solutions $B_{ab}^{ij}=$ $\pm\theta_{a}^{[i}\theta
_{b}^{j]}$ and $A\ $the Levi-Civita connection for a complex metric
$g_{ab}=\theta_{a}\bullet\theta_{b}$. This results in a reduced action which
is a function of the metric only,
\end{subequations}
\[
S(g_{ab})=\mp\int d^{4}xbF,
\]
where $F$ is the scalar curvature $F_{ab}^{ab},$and $b^{2}=\det(g_{ab})$. This
is simply the Einstein-Hilbert action for $SO(4,C)$ general relativity.

The solutions $\pm\frac{1}{2}\epsilon_{ab}^{cd}B_{cd}$ do not correspond to
general relativity \cite{LFKK}, \cite{MPR1}. If $B_{ab}^{ij}=$ $\pm\frac{1}%
{2}\epsilon_{ab}^{cd}B_{cd}$, we obtain a new reduced action,%
\[
S(\theta)=\mp\operatorname{Re}\int d^{4}x\epsilon^{abcd}F_{abcd},
\]
which is zero because of the Bianchi identity $\epsilon^{abcd}F_{abcd}=0$. So
there is no other field equation other than the Bianchi identities.

\section{Plebanski theory with Reality Constraint}

\subsection{Reality Constraint for $b\neq0$}

Let the bivector $2$-form field $B_{ab}^{ij}=\pm\theta_{a}^{[i}\theta_{b}%
^{j]}$ and the space-time metric $g_{ab}=\delta_{ij}\theta_{a}^{i}\theta
_{b}^{j}$. Then, the area metric \cite{MPR1} is defined by
\begin{subequations}
\begin{align}
A_{abcd}  &  =B_{ab}\bullet B_{cd}\label{area1}\\
&  =\frac{1}{2}\eta_{ik}\eta_{jl}B_{ab}^{ij}B_{cd}^{kl}\label{area2}\\
&  =g_{a[c}g_{d]b}. \label{area3}%
\end{align}
Consider an infinitesimal triangle with two sides as real coordinate vectors
$X^{a}$ and $Y^{b}$. Its area $A$ can be calculated in terms of the coordinate
bivector $Q^{ab}=\frac{1}{2}X^{[a}Y^{b]}$ as follows%
\end{subequations}
\[
A^{2}=A_{abcd}Q^{ab}Q^{cd}.
\]
In general $A_{abcd}$ defines a metric on coordinate bivector fields:$<\alpha
,\beta>=A_{abcd}\alpha^{ab}\beta^{cd}$ where $\alpha^{ab}$ and $\beta^{cd}$
are arbitrary bivector fields.

Consider a bivector $2$-form field $B_{ab}^{ij}=$ $\pm\theta_{a}^{[i}%
\theta_{b}^{j]}$ on the real manifold $M$ defined in the last section. Let
$\theta_{a}^{i}$ be non-degenerate complex tetrads. Let $g_{ab}=g_{ab}%
^{R}+ig_{ab}^{I}$, where $g_{ab}^{R}$ and $g_{ab}^{I}$ are the real and the
imaginary parts of $g_{ab}=\theta_{a}\bullet\theta_{b}$.

\begin{theorem}
The area metric being real%
\begin{equation}
\operatorname{Im}(A_{abcd})=0, \label{Reality}%
\end{equation}
is the necessary and the sufficient condition for the non-degenerate metric to
be real or imaginary.
\end{theorem}

\begin{proof}
Equation (\ref{Reality}) is equivalent to the following:
\begin{equation}
g_{ac}^{R}g_{db}^{I}=g_{ad}^{R}g_{cb}^{I}. \label{Reality2.2}%
\end{equation}
From equation (\ref{Reality2.2}) the necessary part of our theorem is
trivially satisfied. Let $g,$ $g^{R}$ and $g^{I}$ be the determinants of
$g_{ab},$ $g_{ab}^{R}$ and $g_{ab}^{I}$ respectively. The consequence of
equation (\ref{Reality2.2}) is that $g=g^{R}+g^{I}$. Since $g\neq0$, one of
$g^{R}$ and $g^{I}$ is non-zero. Let me assume $g^{R}\neq$ $0$ and $g_{R}%
^{ac}$ is the inverse of $g_{ab}^{R}$. Let me multiply both the sides of
equation (\ref{Reality2.2}) by $g_{R}^{ac}$ and sum on the repeated indices.
We get $4g_{db}^{I}=g_{db}^{I}$, which implies $g_{db}^{I}$ $=0.$ Similarly
we\ can show that $g^{I}\neq$ $0$ implies $g_{db}^{R}=0$. So we\ have shown
that the metric is either real or imaginary iff the area metric is real.
\end{proof}

\textit{Since an imaginary metric essentially defines a real geometry,
we\ have shown that the area metric being real is the necessary and the
sufficient condition for } \textit{real geometry (non-degenerate) on the real
manifold }$M$. In one of the appendix I discuss this for any dimensions and
rank of the space-time metric.

To understand the nature of the four volume after imposing the area metric
reality constraint, consider the determinant of both the sides of the equation
$g_{ab}=\theta_{a}\bullet\theta_{b}$,%
\[
g=b^{2},
\]
where $b=\frac{1}{4!}\epsilon^{abcd}B_{ab}\wedge B_{cd}\neq0$. From this
equation we can deduce that $b$ is not sensitive to the fact that the metric
is real or imaginary. But $b$ is imaginary if the metric is Lorentzian
(signature $+++-$ or $---+$) and it is real if the metric is Riemannian or
Kleinien ($++++,----,--++$).

The signature of the metric is directly related to the signature of the area
metric $A_{abcd}=g_{a[c}g_{d]b}$. It can be easily shown that for Riemannian,
Kleinien and Lorentzian geometries the signatures type of $A_{abcd}$ are
$(6,0)$, $(4,2)$ and $(3,3)$ respectively.

Consider the Levi-Civita connection
\[
\Gamma_{bc}^{a}=\frac{1}{2}g^{ad}[\partial_{b}g_{cd}+\partial_{c}%
g_{db}-\partial_{d}g_{bc}]
\]
defined in terms of the metric. From the expression for the connection we can
clearly see that it is real even if the metric is imaginary. Similarly the
Riemann curvature tensor
\[
F_{bcd}^{a}=\partial_{\lbrack c}\Gamma_{d]b}^{a}+\Gamma_{b[c}^{e}\Gamma
_{d]e}^{a}%
\]
is real since it is a function of $\Gamma_{bc}^{a}$ only. But $F_{bc}%
^{ad}=g^{de}F_{bce}^{a}$ and the scalar curvature are real or imaginary
depending on the metric.

In background independent quantum general relativity models, areas are
fundamental physical quantities. In fact the area metric contains the full
information about the metric up to a sign \footnote{For example, please see
the proof of theorem 1 of Ref:\cite{TDRKJS}.}. If $B_{ab}^{R}$ and $B_{ab}%
^{L}$ (vectorial indices suppressed) are the self-dual and the anti-self dual
parts of an arbitrary $B_{ab}^{ij}$, one can calculate the left and right area
metrics as
\[
A_{abcd}^{L}=B_{ab}^{L}\bullet B_{cd}^{L}-\frac{1}{4!}\epsilon^{efgh}%
B_{ef}^{L}\bullet B_{gh}^{L}\epsilon_{abcd}%
\]
and%
\[
A_{abcd}^{R}=B_{ab}^{R}\bullet B_{cd}^{R}+\frac{1}{4!}\epsilon^{efgh}%
B_{ef}^{R}\bullet B_{gh}^{R}\epsilon_{abcd}%
\]
respectively \cite{MPR1}. These metrics are pseudo-scalar component free.
Reisenberger has derived Riemannian general relativity by imposing the
constraint that the left and right area metrics be equal to each other
\cite{MPR1}. This constraint is equivalent to the Plebanski constraint
$B_{ab}\wedge B_{cd}-b\epsilon_{abcd}=0$. I would like to take this one step
further by utilizing the area metric to impose reality constraints on
$SO(4,C)\ $general relativity.

\subsection{Plebanski Action with the reality constraint.}

Next, I\ would like to proceed to modify $SO(4,C)$ general relativity actions
defined before to incorporate the area metric reality constraint. The new
actions are defined as follows:
\begin{equation}
S_{c}(A,B,\bar{B},\phi,q)=\int_{M}\varepsilon^{abcd}B_{ab}\wedge F_{cd}%
d^{4}x+C_{S}+C_{R}, \label{Realaction}%
\end{equation}
and
\[
S_{r}(A,B,\bar{A},\bar{B},\phi,\bar{\phi},q)=\operatorname{Re}S(A,B,\bar
{B},\phi,q),
\]
where%
\begin{equation}
C_{S}=\int_{M_{r}}\frac{b}{2}\phi^{abcd}B_{ab}\wedge B_{cd}d^{4}x \label{CS}%
\end{equation}
and%
\begin{equation}
C_{R}=\int_{M}\frac{\left\vert b\right\vert }{2}q^{abcd}\operatorname{Im}%
\left(  B_{ab}\bullet B_{cd}\right)  d^{4}x. \label{CR}%
\end{equation}
The field $\phi^{abcd}$ is the same as in the last section. The field
$q^{abcd}$ is real with the symmetries of the Riemann curvature tensor. The
$C_{R}$ is the Lagrange multiplier term introduced to impose the area metric
reality constraint.

The field equations corresponding to the extrema of the actions under the $A$
and $\phi$ variations are the same as given in section two. They impose the
condition $B_{ab}^{ij}=$ $\pm\theta_{a}^{[i}\theta_{b}^{j]}$ or $\pm\ast
\theta_{a}^{[i}\theta_{b}^{j]}$ and\ $A$ be the Levi-Civita connection for the
complex metric. The field equations corresponding to the extrema of the
actions under the $q^{abcd}$ variations are $\operatorname{Im}(B_{ab}\bullet
B_{cd})=0$. This, as we discussed before, imposes the condition that the
metric $g_{ab}=\theta_{a}\bullet\theta_{b}$ be real or imaginary\footnote{Also
for $B_{ab}^{ij}=\pm\ast\theta_{a}^{[i}\theta_{b}^{j]},$ it can be verified
that the reality constraint implies that the metric $g_{ab}=\theta_{a}%
\bullet\theta_{b}$ be real or imaginary.}.

Let me assume I have solved the simplicity constraint, the reality constraint
and $dB=0$. Substitute the solutions $B_{ab}^{ij}=$ $\pm\theta_{a}^{[i}%
\theta_{b}^{j]}$ and $A\ $the Levi-Civita connection for a real or imaginary
metric $g_{ab}=\theta_{a}\bullet\theta_{b}$ in the action $S$ . This results
in a reduced action which is a function of the tetrad $\theta_{a}^{i}$ only,
\[
S(\theta)=\mp\operatorname{Re}\int d^{4}xbF.
\]
where $F$ is the scalar curvature $F_{ab}^{ab}$. Recall that $F$ is real or
imaginary depending on the metric. This action reduces to Einstein-Hilbert
action if both the metric and space-time density are simultaneously real or
imaginary. If not, it is zero and there is no field equation involving the
curvature $F_{cd}^{ab}$ tensor other than the Bianchi identities.

If $B_{ab}^{ij}=\pm\ast\theta_{a}^{[i}\theta_{b}^{j]}$ , we get a new reduced
action,%
\begin{equation}
S(\theta)=\mp\operatorname{Re}\int d^{4}x\epsilon^{abcd}F_{abcd},
\label{Non-Einstein}%
\end{equation}
which is zero because of the Bianchi identity $\epsilon^{abcd}F_{abcd}=0$. So
there is no other field equation other than the Bianchi identities.

\section{Discretization}

\subsection{BF theory}

Consider that a continuum manifold is triangulated with four simplices. The
discrete equivalent of a bivector two-form field is the assignment of a
bivector $B_{b}^{ij}$ to each triangle $b$ of the triangulation. Also the
equivalent of a connection one-form is the assignment of a parallel propagator
$g_{eij}$ to each tetrahedron $e$. Using the bivectors and parallel
propagators assigned to the simplices, the actions for general relativity and
BF theory can be rewritten in a discrete form \cite{FoamDer}. The real
$SO(4,C)$ BF action can be discretized as follows \cite{ooguriBFderv}:%
\begin{equation}
S(B_{b},g_{e})=\operatorname{Re}\sum_{b}B_{b}^{ij}ln{H_{bij}.}
\label{eq.bf.des}%
\end{equation}
The $H_{b}$ is the holonomy associated to the triangle $b$. It will be
quantized to get an spin foam model later as done by Ooguri in section five.

\subsection{Barrett--Crane Constraints}

The bivectors $B_{i}$ associated with the ten triangles of a four simplex in a
flat Riemannian space satisfy the following properties called the
Barrett-Crane constraints \cite{BCReimmanion}:

\begin{enumerate}
\item The bivector changes sign if the orientation of the triangle is changed.

\item Each bivector is simple.

\item If two triangles share a common edge, then the sum of the bivectors is
also simple.

\item The sum of the bivectors corresponding to the edges of any tetrahedron
is zero. This sum is calculated taking into account the orientations of the
bivectors with respect to the tetrahedron.

\item The six bivectors of a four simplex sharing the same vertex are linearly independent.

\item The volume of a tetrahedron calculated from the bivectors is real and non-zero.
\end{enumerate}

The items two and three can be summarized as follows:
\[
B_{i}\wedge B_{j}=0~\forall i,j,
\]
where $A\wedge B=\varepsilon_{IJKL}A^{IJ}B^{KL}$ and the $i,j$ represents the
triangles of a tetrahedron. If $i=j$, it is referred to as the simplicity
constraint. If $i\neq j$ it is referred as the cross-simplicity constraints.

Barrett and Crane have shown that these constraints are sufficient to restrict
a general set of ten bivectors $E_{b}$ so that they correspond to the
triangles of a geometric four simplex up to translations and rotations in a
four dimensional flat Riemannian space \cite{BCReimmanion}.

The Barrett-Crane constraints theory can be easily extended to the $SO(4,C)$
general relativity. In this case the bivectors are complex and so the volume
calculated for the sixth constraint is complex. So we need to relax the
condition of the reality of the volume.

We would like to combine the area metric reality constraint with the
Barrett-Crane Constraints. For this we must find the discrete equivalent of
the area metric reality condition. For this let me next discuss the area
metric reality condition in the context of three simplices and four simplices.
I would like to show that the discretized area metric reality constraint
combined with the Barrett-Constraint constraint requires the complex bivectors
associated to a three or four simplex to describe real flat geometries.

\subsubsection{Tetrahedron}

Consider a tetrahedron $t$. Let the numbers $0$ to $3$ denote the vertices of
the tetrahedron. Let me choose the $0$ as the origin of the tetrahedron. Let
$B_{ij}$ be the complex bivector associated with the triangle $0ij$ where $i$
and $j$ denote one of the vertices other than the origin and $i<$ $j$. Let
$B_{0}$ be the complex bivector associated with the triangle $123$. Then
similar to Riemannian general relativity \cite{BCReimmanion}, the
Barrett-Crane constraints\footnote{We do not require to use the fifth
Barrett-Crane constraint since we are only considering one tetrahedron of a
four simplex.} for $SO(4,C)$ general relativity imply that
\begin{subequations}
\label{eq.BC}%
\begin{align}
B_{ij}  &  =a_{i}\wedge a_{j},\label{BC1}\\
B_{0}  &  =-B_{12}-B_{23}-B_{34}, \label{BC2}%
\end{align}
where $a_{i}$, $i=1$ to $3$ are linearly independent complex four vectors
associated to the links $0i$ of the three simplex. Let me choose the vectors
$a_{i}$, $i=1$ to $3$ to be the complex vector basis inside the tetrahedron.
Then the complex $3D$ metric inside the tetrahedron is
\end{subequations}
\begin{equation}
g_{ij}=a_{i}\cdot a_{j}, \label{eq.metric}%
\end{equation}
where the dot is the scalar product on the vectors. This describes a flat
complex three dimensional geometry inside the tetrahedron. The area metric is
given by
\[
A_{ijkl}=g_{i[k}g_{l]j}.
\]
The coordinates of the vectors $a_{i}$ are simply%
\begin{align*}
a_{1}  &  =(1,0,0),\\
a_{2}  &  =(0,1,0),\\
a_{3}  &  =(0,0,1).
\end{align*}
Because of this all of the six possible scalar products made out of the
bivectors $B_{ij}$ are simply the elements of the area metric. From the
discussion of the last section the reality of the area metric simply requires
that the metric $g_{ij}$ be real or imaginary. Since $B_{0}$ is also defined
by equation (\ref{BC2}) its inner product with itself and other bivectors are
real. Thus in the context of a three simplex, the discrete equivalent of the
area metric reality constraint is that the all possible scalar products of
bivectors associated with the triangles of a three simplex be real.

\subsubsection{Four Simplex}

In the case of a four simplex $s$ there are six bivectors $B_{ij}$. There are
four $B_{0}$ type bivectors. Let $B_{i}$ denote the bivector associated to the
triangle made by connecting the vertices other than the origin and vertex $i$.
The Barrett-Crane constraints imply equation (\ref{BC1}) with $i,j=$ $1$ to
$4$. There is one equation for each $B_{i}$ similar to equation (\ref{BC2}).
Now the metric $g_{ij}=a_{i}\cdot a_{j}$ describes a complex four dimensional
flat geometry inside the four simplex $s$. Now assuming we are dealing with
non-degenerate geometry, the reality of the geometry requires the reality of
the area metric. Similar to the three dimensional case, the components of the
area metric are all of the possible scalar products made out of the bivectors
$B_{ij}$. The scalar products of the bivectors $B_{i}$ among themselves or
with $B_{ij}$'s are simple real linear combinations of the scalar products
made from $B_{ij}$'s. So one can propose that the discrete equivalent of the
area metric reality constraint is simply the condition that the scalar product
of these bivectors be real. Let me refer to the later condition as the
bivector scalar product reality constraint.

\begin{theorem}
The necessary and sufficient conditions for a four simplex with real
non-degenerate flat geometry are 1) The $SO(4,C)$ Barrett-Crane
constraints\footnote{The $SO(4,C)$ Barrett-Crane constraints differ from the
real Barrett-Crane constraints by the following:
\par
\begin{enumerate}
\item The bivectors are complex, and
\par
\item The condition for the reality of the volume of tetrahedron is not
required.
\end{enumerate}
} and 2) The reality of all possible bivector scalar products.
\end{theorem}

\begin{proof}
The necessary condition can be shown to be true by straight forward
generalization of the arguments given by Barrett and Crane \cite{BCReimmanion}
and application of the discussions in the last paragraph. The sufficiency of
the conditions follow from the discussion in the last paragraph.
\end{proof}

\section{Spin foam of the $SO(4,C)$ BF model}

Consider a four dimensional submanifold $M$. Let $A$ be a $SO(4,C)$ connection
1-form and $B^{ij}$ a complex bivector valued $2$-form on $M$. Let $F$ be the
curvature 2-form of the connection $A$. Then the real continuum BF\ theory
action defined in section two is,%
\begin{equation}
S_{BF}(A,B_{ij},\bar{A},\bar{B}_{ij})=\operatorname{Re}\int_{M}B\wedge F,
\label{BFaction}%
\end{equation}
where $A,B_{ij}$ and their complex conjugates are considered as independent
free variables. This classical theory is a topological field theory. This
property also holds on spin foam quantization as will be discussed below.

The Spin foam model for the $SO(4,C)$ BF\ theory action can be derived
from\ the discretized BF action by using the path integral quantization as
illustrated in Ref:\cite{ooguriBFderv} for compact groups. Let $\Delta$ be a
simplicial manifold obtained by a triangulation of $M$. Let $g_{e}\in SO(4,C)$
be the parallel propagators associated with the edges (three-simplices)
representing the discretized connection. Let $H_{b}=%
%TCIMACRO{\tprod _{e\supset b}}%
%BeginExpansion
{\textstyle\prod_{e\supset b}}
%EndExpansion
g_{e}$ be the holonomies around the bones (two-simplices) in the four
dimensional matrix representation of $SO(4,C)$ representing the curvature. Let
$B_{b}$ be the $4\times4$ antisymmetric complex matrices corresponding to the
dual Lie algebra of $SO(4,C)$ corresponding to the discrete analog of the $B$
field. Then the discrete BF action is
\[
S_{d}=\operatorname{Re}\sum_{b\in M}tr(B_{b}\ln H_{b}),
\]
which is considered as a function of the $B_{b}$'s and $g_{e}$'s. Here $B_{b}$
the discrete analog of the $B$ field are $4\times4$ antisymmetric complex
matrices corresponding to dual Lie algebra of $SO(4,C)$. The $\ln$ maps from
the group space to the Lie algebra space. The trace is taken over the Lie
algebra indices. Then the\ quantum partition function can be calculated using
the path integral formulation as,%

\[
Z_{BF}(\Delta)=\int\prod_{b}dB_{b}d\bar{B}_{b}\exp(iS_{d})\prod_{e}dg_{e}%
\]%
\begin{equation}
=\int\prod_{b}\delta(H_{b})\prod_{e}dg_{e}, \label{eq.del}%
\end{equation}
where $dg_{e}$ is the invariant measure on the group $SO(4,C)$. The invariant
measure can be defined as the product of the bi-invariant measures on the left
and the right $SL(2,C)$ matrix components. Please see appendix A and B\ for
more details. Similar to the integral measure on the $B$'s an explicit
expression for the $dg_{e}$ involves product of conjugate measures of complex coordinates.

Now consider the identity
\begin{equation}
\delta(g)=\frac{1}{64\pi^{8}}\int d_{\omega}tr(T_{\omega}(g))d\omega,
\label{eq.del.exp}%
\end{equation}
where the $T_{\omega}(g)$ is a unitary representation of $SO(4,C),$ where
$\omega=$ $\left(  \chi_{L},\chi_{R}\right)  $ such that $n_{L}+n_{R}$ is
even, $d_{\omega}=\left\vert \chi_{L}\chi_{R}\right\vert ^{2}$. The details of
the representation theory is discussed in appendix B. The integration with
respect to $d\omega$ in the above equation is interpreted as the summation
over the discrete $n$'s and the integration over the continuous $\rho$'s.

By substituting the harmonic expansion for $\delta(g)$ into the equation
(\ref{eq.del}) we can derive the spin foam partition of the $SO(4,C)\ BF$
theory as explained in Ref:\cite{JCBintro} or Ref:\cite{ooguriBFderv}. The
partition function is defined using the $SO(4,C)$ intertwiners and the
$\{15\omega\}$ symbols.

The relevant intertwiner for the $BF\ $spin foam is%
\[
i_{e}=%
%TCIMACRO{\FRAME{itbphF}{0.518in}{0.7628in}{0.3814in}{}{}{4dintw.eps}%
%{\special{ language "Scientific Word";  type "GRAPHIC";
%maintain-aspect-ratio TRUE;  display "USEDEF";  valid_file "F";
%width 0.518in;  height 0.7628in;  depth 0.3814in;  original-width 2.6913in;
%original-height 3.9773in;  cropleft "0";  croptop "1";  cropright "1";
%cropbottom "0";  filename '4dintw.eps';file-properties "XNPEU";}}}%
%BeginExpansion
\raisebox{-0.3814in}{\includegraphics[
height=0.7628in,
width=0.518in
]%
{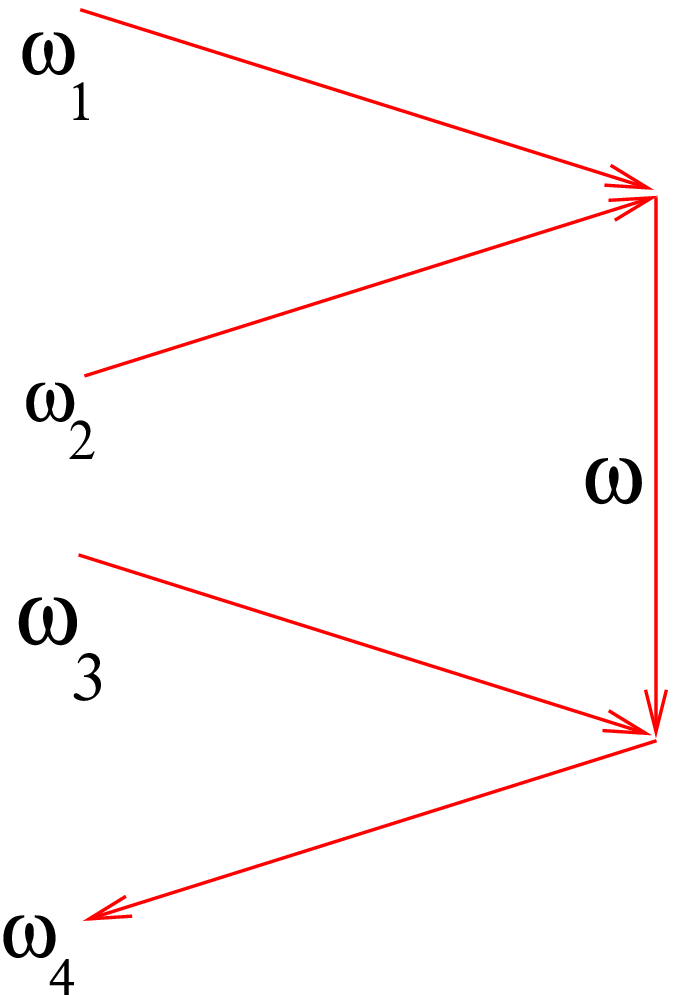}%
}%
%EndExpansion
.
\]
The nodes where the three links meet are the Clebsch-Gordan coefficients of
$SO(4,C)$. The Clebsch-Gordan coefficients of $SO(4,C)$ are just the product
of the Clebsch-Gordan coefficients of the left and the right handed $SL(2,C)$
components. The Clebsch-Gordan coefficients of $SL(2,C)$ are discussed in the
references \cite{IMG} and \cite{NaimarckClebsch}.

The quantum amplitude associated with each simplex $s$ is given below and can
be referred to as the $\{15\omega\}$ symbol,%

\[
\{15\omega\}=%
%TCIMACRO{\FRAME{itbphF}{1.6016in}{1.548in}{0.8026in}{}{}{15w.eps}%
%{\special{ language "Scientific Word";  type "GRAPHIC";
%maintain-aspect-ratio TRUE;  display "USEDEF";  valid_file "F";
%width 1.6016in;  height 1.548in;  depth 0.8026in;  original-width 5.5011in;
%original-height 5.316in;  cropleft "0";  croptop "1";  cropright "1";
%cropbottom "0";  filename '15w.eps';file-properties "XNPEU";}}}%
%BeginExpansion
\raisebox{-0.8026in}{\includegraphics[
height=1.548in,
width=1.6016in
]%
{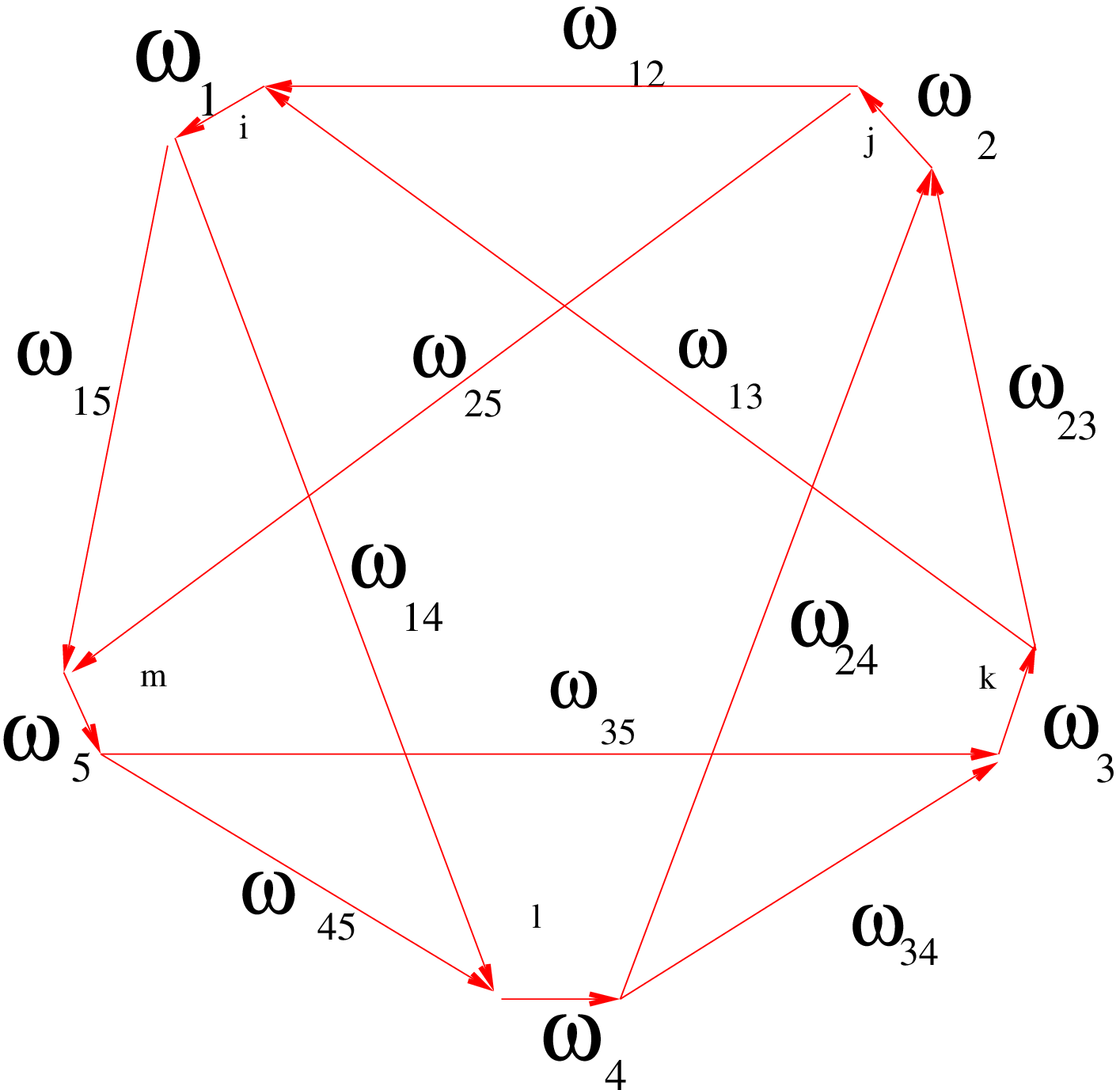}%
}%
%EndExpansion
.
\]
The final partition function is
\begin{equation}
Z_{BF}(\Delta)=\int\limits_{\{\omega_{b,}\omega_{e}\}}\prod_{b}\frac
{d_{\omega_{b}}}{64\pi^{8}}\prod_{s}Z_{BF}(s)\prod_{b}d\omega_{b}\prod
_{e}d\omega_{e}, \label{eq.6}%
\end{equation}
where the $Z_{BF}(s)=\{15\omega\}$ is the amplitude for a four-simplex $s$.
The $d_{\omega_{b}}=\left\vert \chi_{L}\chi_{R}\right\vert ^{2}$ term is the
quantum amplitude associated with the bone $b$.\ Here $\omega_{e}$ is the
internal representation used to define the intertwiners. Usually $\omega_{e}$
is replaced by $i_{e}$ to indicate the intertwiner. The set $\left\{
\boldsymbol{\omega}_{b,}\boldsymbol{\omega}_{e}\right\}  $ of all $\omega_{b}%
$'s and $\omega_{e}$'s is usually called a \textbf{coloring} of the bones and
the edges. This partition function may not be finite in general.

It is well known that the$\ BF$ theories are topological field theories. A
priori one cannot expect this to be true for the case of the BF spin foam
models because of the discretization of the BF action. For the spin foam
models of the BF\ theories for compact groups, it has been shown that the
partition functions are triangulation independent up to a factor
\cite{SpinFoamDiag}. This analysis is purely based on spin foam diagrammatics
and is independent of the group used as long the BF spin foam is defined
formally by equation (\ref{eq.del}) and the harmonic expansion in equation
(\ref{eq.del.exp}) is formally valid. So one can apply the spin foam
diagrammatics analysis directly to the $SO(4,C)$ BF spin foam and write down
the triangulation independent partition function as%

\[
Z_{BF}^{^{\prime}}(\Delta)=\tau^{n_{4}-n_{3}}Z_{BF}(\Delta)
\]
using the result from \cite{SpinFoamDiag}. In the above equation $n_{4},n_{3}$
is number of four bubbles and three bubbles in the triangulation $\Delta$ and
\begin{align*}
\tau &  =\delta_{SO(4,C)}(I)\\
&  =\frac{1}{64\pi^{8}}\int d_{\omega}^{2}d\omega.
\end{align*}
The above integral is divergent and so the partition functions need not be
finite. The normalized partition function is to be considered as the proper
partition function because the BF\ theory is supposed to be topological and so
triangulation independent.

\section{The $SO(4,C)$ Barrett-Crane Model}

\subsection{Introduction}

My goal here is to systematically construct the Barrett-Crane model of the
$SO(4,C)$ general relativity. In the previous section I\ discussed the
$SO(4,C)$ BF spin foam model. The basic elements of the BF spin foams are spin
networks built on graphs dual to the triangulations of the four simplices with
arbitrary intertwiners and the principal unitary representations of $SO(4,C)$
discussed in appendix B. These closed spin networks can be considered as
quantum states of four simplices in the BF\ theory and the essence of these
spin networks is mainly gauge invariance. To construct a spin foam model of
general relativity these spin networks need to be modified to include the
Plebanski Constraints in the discrete form.

A\ quantization of a four-simplex for the Riemannian general relativity was
proposed by Barrett and Crane \cite{BCReimmanion}. The bivectors $B_{i}$
associated with the ten triangles of a four-simplex in a flat Riemannian space
satisfy the properties called the Barrett-Crane constraints\footnote{I\ would
like to refer the readers to the original paper \cite{BCReimmanion} for more
details.}. They have been listed in section 4.2 which are repeated below for convenience:

\begin{enumerate}
\item The bivector changes sign if the orientation of the triangle is changed.

\item Each bivector is simple.

\item If two triangles share a common edge, then the sum of the bivectors is
also simple.

\item The sum of the bivectors corresponding to the edges of any tetrahedron
is zero. This sum is calculated taking into account the orientations of the
bivectors with respect to the tetrahedron.

\item The six bivectors of a four-simplex sharing the same vertex are linearly independent.

\item The volume of a tetrahedron calculated from the bivectors is real and non-zero.
\end{enumerate}

The items two and three can be summarized as follows:
\[
B_{i}\wedge B_{j}=0~\forall i,j,
\]
where $A\wedge B=\varepsilon_{IJKL}A^{IJ}B^{KL}$ and the $i,j$ represents the
triangles of a tetrahedron. If $i=j$, it is referred to as the simplicity
constraint. If $i\neq j$ it is referred as the cross-simplicity constraints.

Barrett and Crane have shown that these constraints are sufficient to restrict
a general set of ten bivectors $E_{b}$ so that they correspond to the
triangles of a geometric four-simplex up to translations and rotations in a
four dimensional flat Riemannian space.

The Barrett-Crane constraints theory can be trivially extended to the
$SO(4,C)$ general relativity. In this case the bivectors are complex and so
the volume calculated for the sixth constraint is complex. So we need to relax
the condition of the reality of the volume.

A quantum four-simplex for Riemannian general relativity is defined by
quantizing the Barrett-Crane constraints \cite{BCReimmanion}. The bivectors
$B_{i}$ are promoted to the Lie operators $\hat{B}_{i}$ on the representation
space of the relevant group and the Barrett-Crane constraints are imposed at
the quantum level. A\ four-simplex has been quantized and studied in the case
of the Riemannian general relativity before \cite{BCReimmanion}. All the first
four constraints have been rigorously implemented in this case. The last two
constraints are inequalities and they are difficult to impose. This could be
related to the fact that the Riemannian Barrett-Crane model reveal the
presence of degenerate sectors \cite{BaezEtalAsym}, \cite{JWBCS} in the
asymptotic limit \cite{JWBRW} of the model. For these reasons here after
I\ would like to refer to a spin foam model that satisfies only the first four
constraints as an \textit{essential Barrett-Crane model}, While a spin foam
model that satisfies all the six constraints as a \textit{rigorous
Barrett-Crane model}.

Here I\ would like to derive the essential $SO(4,C)$ Barrett-Crane model. For
this one must deal with complex bivectors instead of real bivectors. The
procedure that I\ would like to use to solve the constraints can be carried
over directly to the Riemannian Barrett-Crane\ model. This derivation
essentially makes the derivation of the Barrett-Crane intertwiners for the
real and the complex Riemannian general relativity more rigorous.

\subsection{The SO(4,C)\ intertwiner}

The group $SO(4,C)$ is locally isomorphic to $\frac{SL(2,C)\times
SL(2,C)}{Z_{2}}$. An element $B$ of the Lie algebra space of $SO(4,C)$ can be
split into the left and the right handed $SL(2,C)$ components,%
\begin{equation}
B=B_{L}+B_{R}.
\end{equation}
There are two Casimir operators for $SO(4,C)$ which are $\varepsilon
_{IJKL}B^{IJ}B^{KL}$ and $\eta_{IK}\eta_{JL}B^{IJ}B^{KL}$, where $\eta_{IK}$
is the flat Euclidean metric. In terms of the left and right handed split I
can expand the Casimir operators as%
\[
\varepsilon_{IJKL}B^{IJ}B^{KL}=B_{L}\cdot B_{L}-B_{R}\cdot B_{R}~~\text{and}%
\]%
\[
\eta_{IK}\eta_{JL}B^{IJ}B^{KL}=B_{L}\cdot B_{L}+B_{R}\cdot B_{R},
\]
where the dot products are the trace in the $SL(2,C)$ Lie algebra coordinates.

The bivectors are to be quantized by promoting the Lie algebra vectors to Lie
operators on the unitary representation space of $SO(4,C)\approx$
$\frac{SL(2,C)\times SL(2,C)}{Z_{2}}$. The relevant unitary representations of
$SO(4,C)\simeq$ $SL(2,C)\otimes SL(2,C)/Z_{2}$ are labeled by a pair
($\chi_{L},$ $\chi_{R}$) such that $n_{L}+n_{R}$ is even (appendix B). The
elements of the representation space $D_{\chi_{L}}\otimes$ $D_{\chi_{R}}$ are
the eigen states of the Casimirs and on them the operators reduce to the
following:
\begin{equation}
\varepsilon_{IJKL}\hat{B}^{IJ}\hat{B}^{KL}=\frac{\chi_{L}^{2}-\chi_{R}^{2}}%
{2}\hat{I}~~\text{and} \label{eq.casimer1}%
\end{equation}%
\begin{equation}
\eta_{IK}\eta_{JL}\hat{B}^{IJ}\hat{B}^{KL}=\frac{\chi_{L}^{2}+\chi_{R}^{2}%
-2}{2}\hat{I}. \label{eq.casimer2}%
\end{equation}
The equation (\ref{eq.casimer1}) implies that on $D_{\chi_{L}}\otimes$
$D_{\chi_{R}}$ the simplicity constraint $B\wedge B=0$ is equivalent to the
condition $\chi_{L}=\pm\chi_{R}$. I would like to find a representation space
on which the representations of $SO(4,C)$ are restricted precisely by
$\chi_{L}$ $=$ $\pm\chi_{R}$. Since a $\chi$ representation is equivalent to
$-\chi$ representations \cite{IMG}, $\chi_{L}=+\chi_{R}$ case is equivalent to
$\chi_{L}=-\chi_{R}$ \cite{IMG}.

The Barrett-Crane intertwiner for Riemannian general relativity has been
systematically quantized in Ref:\cite{MyRigorousSpinFoam}. Since the
representation theory for $SO(4,C)$ is similar to that of $SO(4,R),$ the
systematic derivation can be generalized to $SO(4,C)$ general
relativity\footnote{Readers can refer to the preprint
Ref:\cite{FoamsComplexReal} for details of derivations of $SO(4,C)$
Barrett-Crane intertwiner.}.

The components of the Barrett-Crane intertwiner $\left\vert \Psi\right\rangle
\in\bigotimes\limits_{i}D_{\chi_{i}}\otimes D_{\chi_{i}}^{\ast}$ can be
written using the Gelfand-Naimarck representation theory \cite{IMG} as:%
\[
\left\vert \Psi\right\rangle =\int\limits_{CS^{3}}%
%TCIMACRO{\FRAME{itbphF}{1.3474in}{1.2955in}{0.6529in}{}{\Qlb{11}%
%}{bcintrep.eps}{\special{ language "Scientific Word";  type "GRAPHIC";
%display "USEDEF";  valid_file "F";  width 1.3474in;  height 1.2955in;
%depth 0.6529in;  original-width 6.4688in;  original-height 6.2457in;
%cropleft "0";  croptop "1";  cropright "1";  cropbottom "0";
%filename 'BCIntRep.eps';file-properties "XNPEU";}}}%
%BeginExpansion
\raisebox{-0.6529in}{\includegraphics[
height=1.2955in,
width=1.3474in
]%
{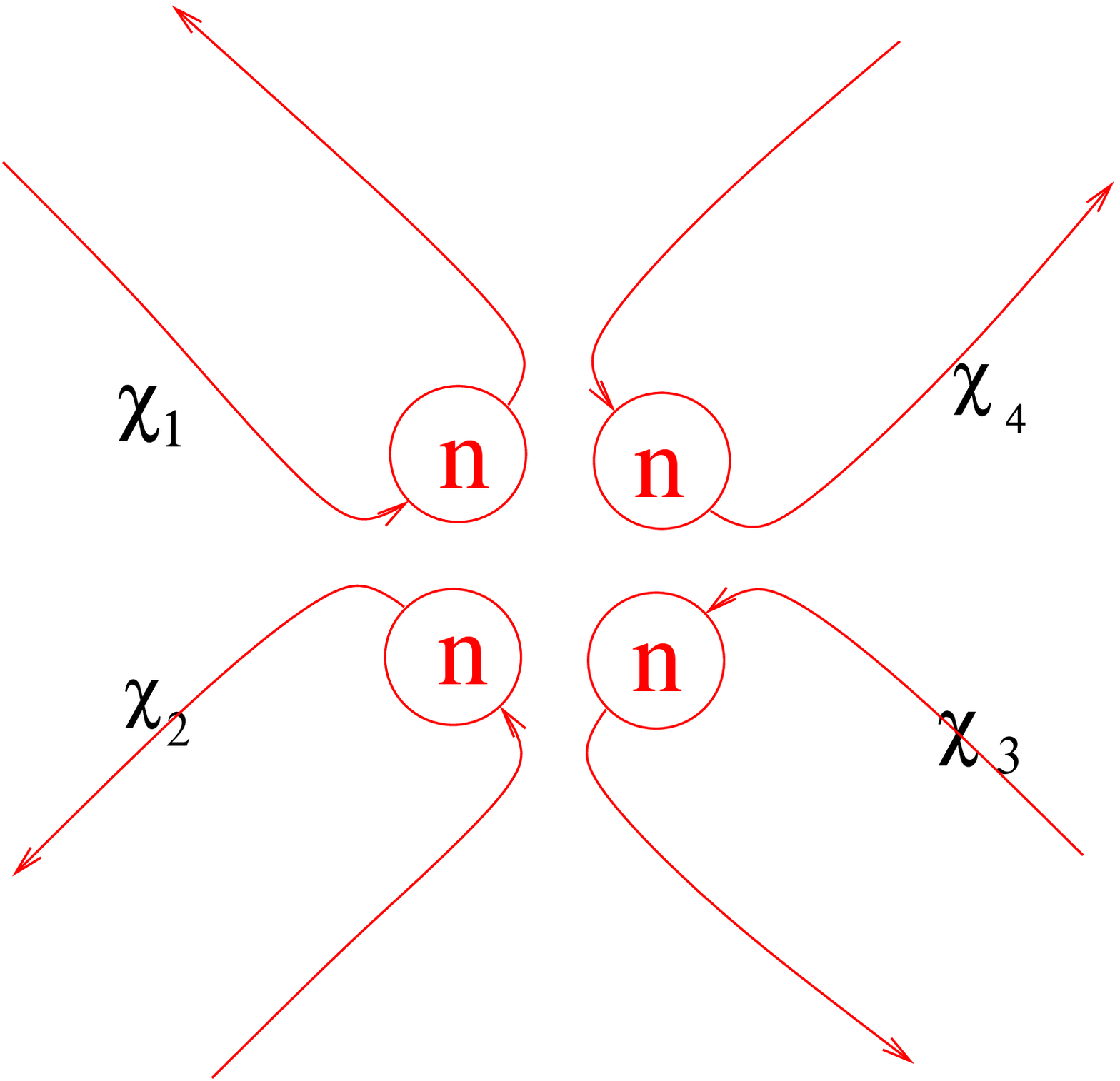}%
}%
%EndExpansion
dn.
\]
Since $SL(2,C)\approx CS^{3},$ using the following graphical identity:%

\[
\int_{SL(2,C)}%
%TCIMACRO{\FRAME{itbphF}{0.9824in}{1.1338in}{0.5708in}{}{\Qlb{12}%
%}{4intertwine.eps}{\special{ language "Scientific Word";  type "GRAPHIC";
%maintain-aspect-ratio TRUE;  display "USEDEF";  valid_file "F";
%width 0.9824in;  height 1.1338in;  depth 0.5708in;  original-width 2.7319in;
%original-height 3.1531in;  cropleft "0";  croptop "1";  cropright "1";
%cropbottom "0";  filename '4intertwine.eps';file-properties "XNPEU";}}}%
%BeginExpansion
\raisebox{-0.5708in}{\includegraphics[
height=1.1338in,
width=0.9824in
]%
{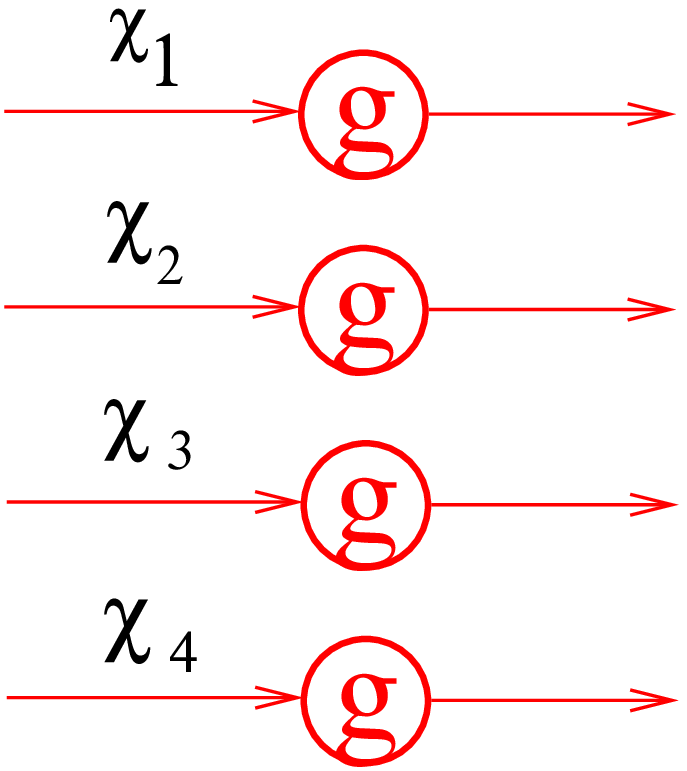}%
}%
%EndExpansion
dg=\int%
%TCIMACRO{\FRAME{itbphF}{2.0877in}{1.1122in}{0.5613in}{}{\Qlb{13}%
%}{edge4int3.eps}{\special{ language "Scientific Word";  type "GRAPHIC";
%maintain-aspect-ratio TRUE;  display "USEDEF";  valid_file "F";
%width 2.0877in;  height 1.1122in;  depth 0.5613in;  original-width 2.5668in;
%original-height 1.3534in;  cropleft "0";  croptop "1";  cropright "1";
%cropbottom "0";  filename 'edge4int3.eps';file-properties "XNPEU";}}}%
%BeginExpansion
\raisebox{-0.5613in}{\includegraphics[
height=1.1122in,
width=2.0877in
]%
{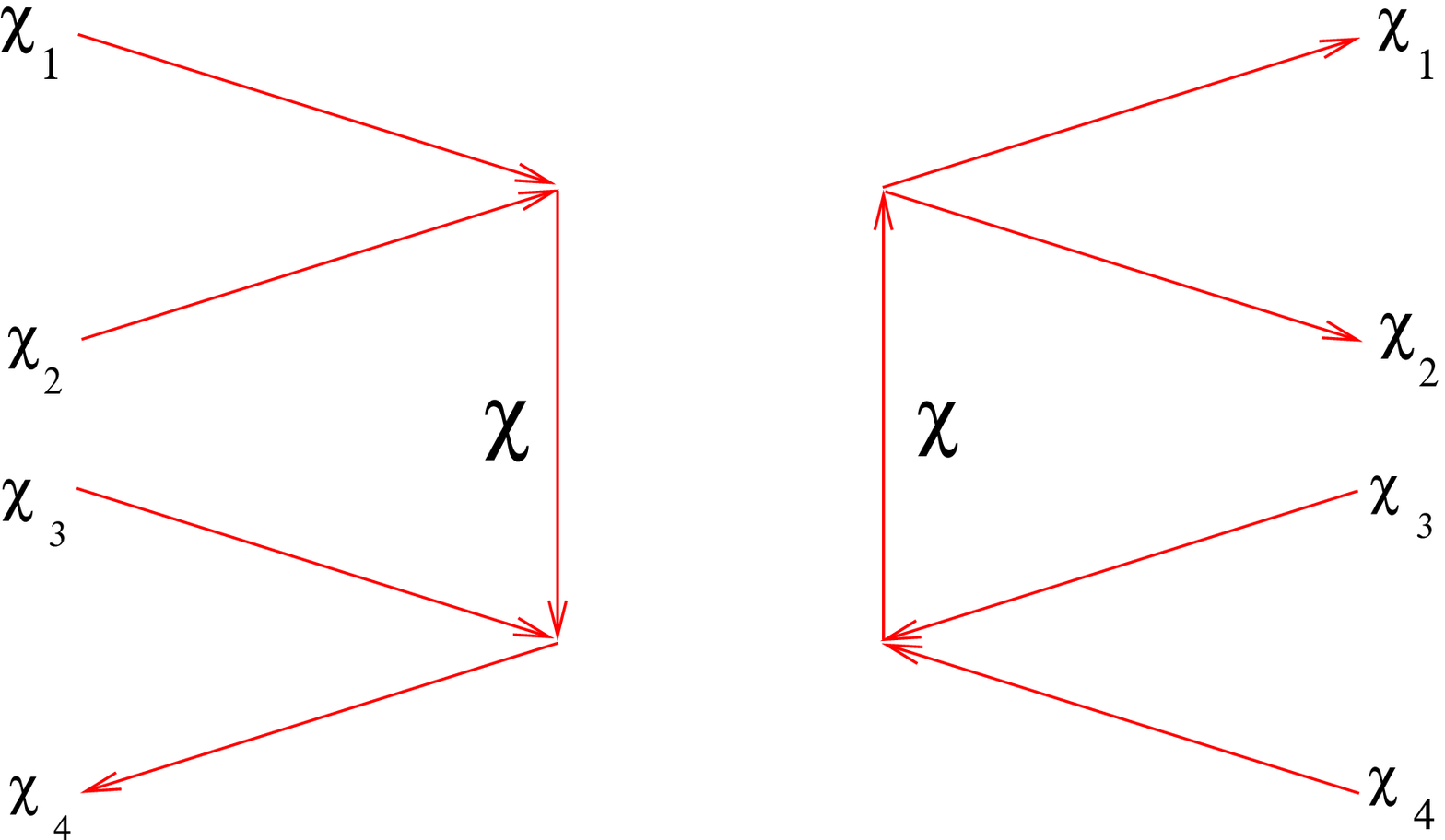}%
}%
%EndExpansion
\frac{8\pi^{4}}{\chi\bar{\chi}}d\chi,
\]
the Barrett-Crane solution can be rewritten as%

\[
\left\vert \Psi\right\rangle =\int%
%TCIMACRO{\FRAME{itbphF}{2.1932in}{0.9617in}{0.4817in}{}{\Qlb{14}}%
%{4xxbc.eps}{\special{ language "Scientific Word";  type "GRAPHIC";
%maintain-aspect-ratio TRUE;  display "USEDEF";  valid_file "F";
%width 2.1932in;  height 0.9617in;  depth 0.4817in;  original-width 5.2088in;
%original-height 2.2684in;  cropleft "0";  croptop "1";  cropright "1";
%cropbottom "0";  filename '4XXBC.eps';file-properties "XNPEU";}}}%
%BeginExpansion
\raisebox{-0.4817in}{\includegraphics[
height=0.9617in,
width=2.1932in
]%
{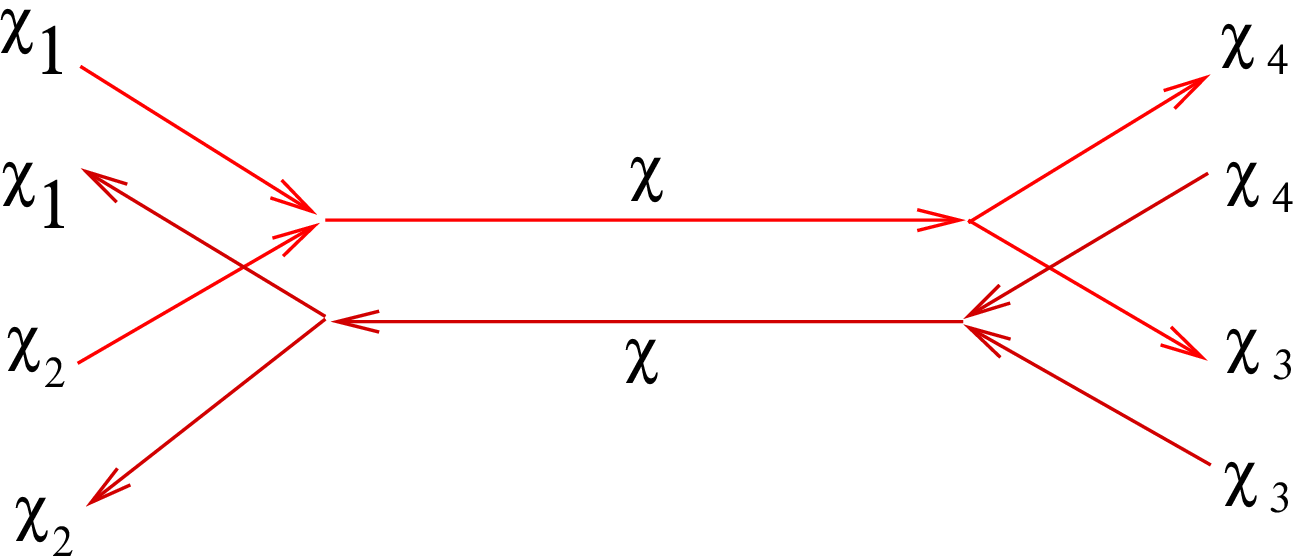}%
}%
%EndExpansion
\frac{8\pi^{4}}{\chi\bar{\chi}}d\chi,
\]
which emerges as an intertwiner in the familiar form in which Barrett and
Crane proposed it for the Riemannian general relativity. It can be clearly
seen that the simple representations for $SO(4,R)$ ($J_{L}=$ $J_{R}$) has been
replaced by the simple representation of $SO(4,C)$ ($\chi_{L}=$ $\pm\chi_{R}$).

All the analysis done until for the $SO(4,C)\ $Barrett-Crane theory can be
directly applied to the Riemannian$\ $Barrett-Crane theory. The
correspondences between the two models are listed in the following
table\footnote{BC stands for Barrett-Crane. For $\chi_{L}$ and $\chi_{R}$ we
have $n_{L}+n_{R}=even$.
\par
{}}:

\begin{center}%
\begin{tabular}
[c]{lll}%
\textbf{Property} & $SO(4,R)$ BC model & $SO(4,C)$ BC model\\
Gauge group & $SO(4,R)\approx\frac{SL(2,C)\otimes SL(2,C)}{Z_{2}}$ &
$SO(4,C)\approx\frac{SU(2)\otimes SU(2)}{Z_{2}}$\\
Representations & $J_{L},J_{R}$ & $\chi_{L},\chi_{R}$\\
Simple representations & $J_{L}=J_{R}$ & $\chi_{L}=\pm\chi_{R}$\\
Homogenous space & $S^{3}\approx SU(2)$ & $CS^{3}\approx SL(2,C)$%
\end{tabular}

\end{center}

\subsection{The Spin Foam Model for the $SO(4,C)$ General Relativity.}

The $SO(4,C)$ Barrett-Crane intertwiner derived in the previous section can be
used to define a $SO(4,C)$ Barrett-Crane spin foam model. The amplitude
$Z_{BC}(s)$ of a four-simplex $s$ is given by the $\{10\chi\}_{SO(4,C)}$
symbol given below:%

\begin{equation}
\{10\chi\}_{SO(4,C)}=%
%TCIMACRO{\FRAME{itbphF}{1.5134in}{1.4131in}{0.7022in}{}{\Qlb{15}}%
%{10xbc.eps}{\special{ language "Scientific Word";  type "GRAPHIC";
%display "USEDEF";  valid_file "F";  width 1.5134in;  height 1.4131in;
%depth 0.7022in;  original-width 5.0254in;  original-height 5.2511in;
%cropleft "0";  croptop "1";  cropright "1";  cropbottom "0";
%filename '10XBC.eps';file-properties "XNPEU";}}}%
%BeginExpansion
\raisebox{-0.7022in}{\includegraphics[
height=1.4131in,
width=1.5134in
]%
{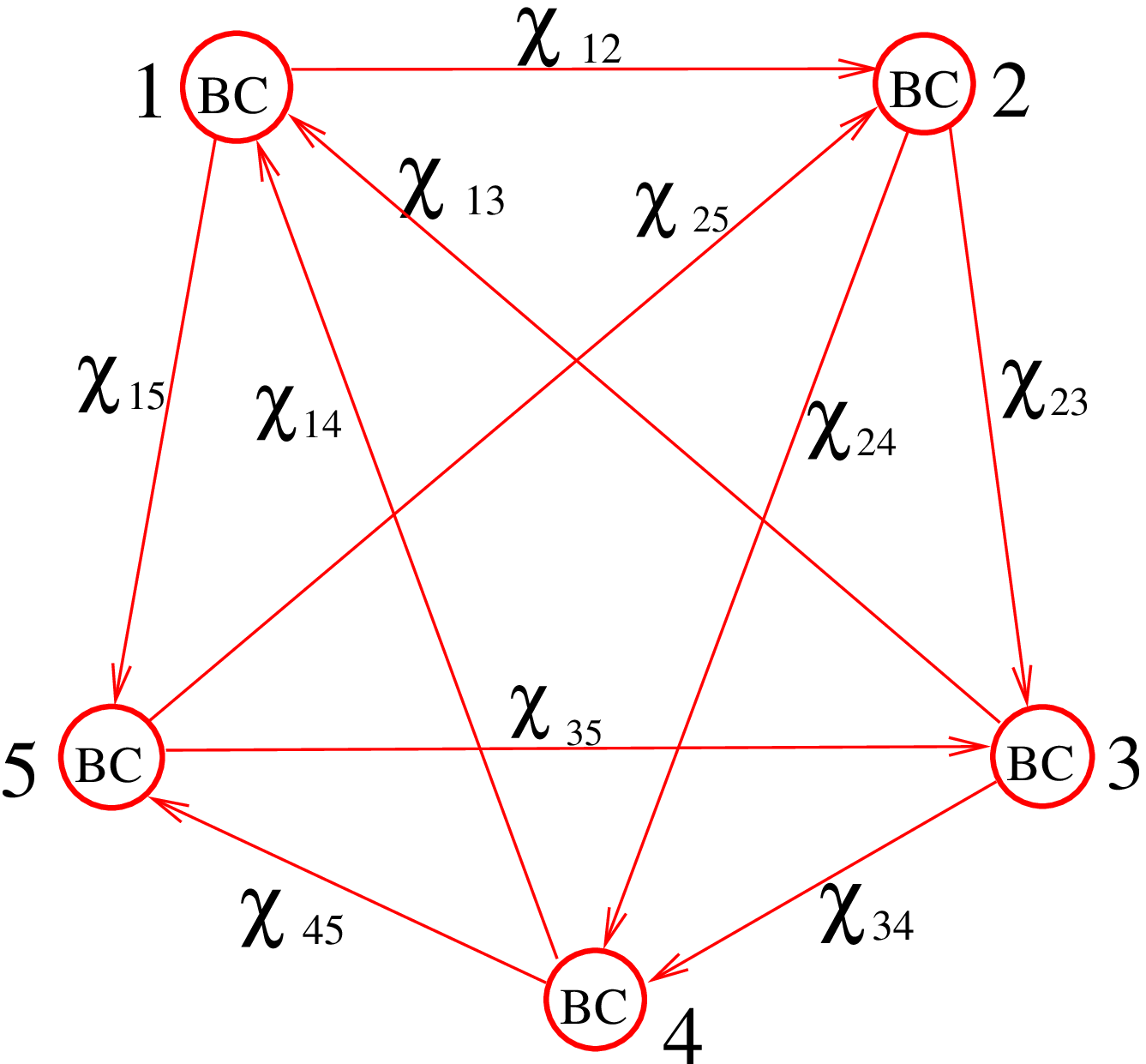}%
}%
%EndExpansion
\text{,} \label{10w}%
\end{equation}
where the circles are the Barrett-Crane intertwiners. The integers represent
the tetrahedra and the pairs of integers represent triangles. The intertwiners
use the four $\chi$'s associated with the links that emerge from it for its
definition in equation (\ref{10w}). In the next subsection, the propagators of
this theory are defined and the $\{10\chi\}$ symbol is expressed in terms of
the propagators in the subsubsection that follows it.

The $SO(4,C)$ Barrett-Crane partition function of the spin foam associated
with the four dimensional simplicial manifold with a triangulation $\Delta$
is
\begin{equation}
Z(\Delta)=\sum_{\left\{  \chi_{b}\right\}  }\left(  \prod_{b}\frac{d_{\chi
_{b}}^{2}}{64\pi^{8}}\right)  \prod_{s}Z(s), \label{PartitionFunction}%
\end{equation}
where $Z(s)$ is the quantum amplitude associated with the $4$-simplex $s$ and
the $d_{\chi_{b}}$ adopted from the spin foam model of the $BF\ $theory can be
interpreted as the quantum amplitude associated with the bone $b$.

\subsection{The Features of the $SO(4,C)$ Spin Foam}

\begin{itemize}
\item Areas: The squares of the areas of the triangles (bones) of the
triangulation are given by $\eta_{IK}\eta_{JL}B^{IJ}B^{KL}$. The eigen values
of the squares of the areas in the $SO(4,C)$ Barrett-Crane model from equation
(\ref{eq.2}) are given by
\begin{align*}
\eta_{IK}\eta_{JL}\hat{B}_{b}^{IJ}\hat{B}_{b}^{KL}  &  =\left(  \chi
^{2}-1\right)  \hat{I}\\
&  =\left(  \frac{n^{2}}{2}-\rho^{2}-1+i\rho n\right)  \hat{I}.
\end{align*}
One can clearly see that the area eigen values are complex. The $SO(4,C)$
Barrett-Crane model relates to the $SO(4,C)$ general relativity. Since in the
$SO(4,C)$ general relativity the bivectors associated with any two dimensional
flat object are complex, it is natural to expect that the areas defined in
such a theory are complex too. This is a generalization of the concept of the
space-like and the time-like areas for the real general relativity models:
Area is imaginary if it is time-like and real if it is space-like.

\item Propagators: Laurent and Freidel have investigated the idea of
expressing simple spin networks as Feynman diagrams \cite{SimpleSpinNetworks}.
Here we will apply this idea to the $SO(4,C)$ simple spin networks. Let
$\Sigma$ be a triangulated three surface. Let $n_{i}\in CS^{3}$ be a vector
associated with the $i^{th}$ tetrahedron of the $\Sigma$. The propagator of
the $SO(4,C)$ Barrett-Crane model associated with the triangle $ij$ is given
by%
\begin{align*}
G_{\chi_{ij}}(n_{i},n_{j})  &  =Tr(T_{\chi_{ij}}(g(n_{i}))T_{\chi_{ij}}^{\dag
}(g(n_{j})))\\
&  =Tr(T_{\chi_{ij}}(g(n_{i})g^{-1}(n_{j}))),
\end{align*}
where $\chi_{ij}$ is a representation associated with the triangle common to
the $i^{th}$ and the $j^{th}$ tetrahedron of $\Sigma$. If $X$ and $Y$ belong
to $CS^{3}$ then
\[
tr\left(  \mathfrak{g}(X)\mathfrak{g}(Y)^{-1}\right)  =2X.Y,
\]
where $X.Y$ is the Euclidean dot product and $tr$ is the matrix trace. If
$\lambda=e^{t}$ and $\frac{1}{\lambda}$ are the eigen values of $g(X)g(Y)^{-1}%
$ then,%
\begin{align*}
\lambda+\lambda^{-1}  &  =2X.Y\\
X.Y  &  =\cosh(t).
\end{align*}
From the expression for the trace of the $SL(2,C)$ unitary representations,
(appendix A, \cite{IMG}) I have the propagator for the $SO(4,C)$ Barrett-Crane
model calculated as%
\[
G_{\chi_{ij}}(n_{i},n_{j})=\dfrac{\cos(\rho_{ij}\eta_{ij}+n_{ij}\theta_{ij}%
)}{\left\vert \sinh(\eta_{ij}+i\theta_{ij})\right\vert ^{2}},
\]
where $\eta_{ij}+i\theta_{ij}$ is defined by $n_{i}.n_{j}=\cosh(\eta
_{ij}+i\theta_{ij})$. Two important properties of the propagators are listed below.

\begin{enumerate}
\item Using the expansion for the delta on $SL(2,C)$ I have
\begin{align*}
\delta_{CS^{3}}(X,Y)  &  =\delta_{SL(2,C)}(g(X)g^{-1}(Y))\\
&  =\frac{1}{8\pi^{4}}\int\bar{\chi}\chi Tr(T_{\chi}(g(X)g^{-1}(Y))d\chi,
\end{align*}
where the suffix on the deltas indicate the space in which it is defined.
Therefore%
\[
\int\bar{\chi}\chi G_{\chi}(X,Y))=8\pi^{4}\delta_{CS^{3}}(X,Y).
\]

\item Consider the orthonormality property of the principal unitary
representations of $SL(2,C)$ given by%
\begin{align*}
&  \int_{CS^{3}}T_{\acute{z}_{1}\chi_{1}}^{z_{1}}(g(X))T_{\acute{z}_{2}%
\chi_{2}}^{\dag z_{2}}(g(X))dX\\
&  =\frac{8\pi^{4}}{\chi_{1}\bar{\chi}_{1}}\delta(\chi_{1}-\chi_{2}%
)\delta(z_{1}-\acute{z}_{1})\delta(z_{2}-\acute{z}_{2}),
\end{align*}
where the delta on the $\chi$'s is defined up to a sign of them$.$ From this I
have%
\[
\int_{CS^{3}}G_{\chi_{1}}(X,Y)G_{\chi_{2}}(Y,Z)dY=\frac{8\pi^{4}}{\chi_{1}%
\bar{\chi}_{1}}\delta(\chi_{1}-\chi_{2})G_{\chi_{1}}(X,Z).
\]

\end{enumerate}

\item The $\{10\chi\}$ symbol can be defined using the propagators on the
complex three sphere as follows:%
\begin{align*}
Z(s)  &  =\int_{x_{k}\in CS^{3}}%
%TCIMACRO{\dprod \limits_{i<j}}%
%BeginExpansion
{\displaystyle\prod\limits_{i<j}}
%EndExpansion
T_{\chi_{ij}}(\mathfrak{g}(x_{i})\mathfrak{g}(x_{j}))%
%TCIMACRO{\dprod \limits_{k}}%
%BeginExpansion
{\displaystyle\prod\limits_{k}}
%EndExpansion
dx_{k},\\
&  =\int_{\forall x_{k}\in CS^{3}}%
%TCIMACRO{\dprod \limits_{i<j}}%
%BeginExpansion
{\displaystyle\prod\limits_{i<j}}
%EndExpansion
G_{\chi_{ij}}(x_{i}\mathfrak{,}x_{j})%
%TCIMACRO{\dprod \limits_{k}}%
%BeginExpansion
{\displaystyle\prod\limits_{k}}
%EndExpansion
dx_{k},
\end{align*}
where $i$ denotes a tetrahedron of the four-simplex. For each tetrahedron
$k,~$a free variable $x_{k}\in CS^{3}$ is associated. For each triangle $ij$
which is the intersection of the $i$'th and the $j$'th tetrahedron, a
representation of $SL(2,C)$ denoted by $\chi_{ij}$ is associated.

\item Discretization Dependence and Local Excitations: It is well known that
the BF theory is discretization independent and is topological. The$\ $spin
foam for the $SO(4,C)$ general relativity is got by imposing the Barrett-Crane
constraints on the $BF\ $Spin foam. After the imposition of the Barrett-Crane
constraints the theory loses the discretization independence and the
topological nature. This can be seen in many ways.

\begin{itemize}
\item The simplest reason is that the $SO(4,C)\ $Barrett-Crane model
corresponds to the quantization of the discrete $SO(4,C)$ general relativity
which has local degrees of freedom.

\item After the restriction of the representations involved in BF spin foams
to the simple representations and the intertwiners to the Barrett-Crane
intertwiners, various important identities used in the spin foam diagrammatics
and proof of the discretization independence of the BF\ theory spin foams in
Ref:\cite{SpinFoamDiag} are no longer available.

\item The BF partition function is simply gauge invariant measure of the
volume of space of flat connections. Consider the following harmonic expansion
of the delta function which was used in the derivation of the $SO(4,C)$ BF
theory:%
\[
\delta(g)=\frac{1}{8\pi^{4}}\int d_{\omega}tr(T_{\omega}(g))d\omega.
\]
Imposition of the Barrett-Crane constraints on the BF theory spin foam,
suppresses the terms corresponding to the non-simple representations. If only
the simple representations are allowed in the right hand side, it is no longer
peaked at the identity. This means that the partition function for
the\ $SO(4,C)$\ Barrett-Crane model involves contributions only from the
non-flat connections which has local information.

\item In the asymptotic limit study of the $SO(4,C)$ spin foams in section
four of Ref:\cite{FoamsComplexReal} the discrete version of the $SO(4,C)$
general relativity (Regge calculus) is obtained. The Regge calculus action is
clearly discretization dependent and non-topological.
\end{itemize}

\item The real Barrett-Crane models that are discussed in the next section are
the restricted form of the $SO(4,C)\ $Barrett-Crane model. The above reasoning
can be applied to argue that they are also discretization dependent.
\end{itemize}

\section{Spin Foams for 4D Real General Relativity and reality constraints}

\subsection{The Formal Structure of Barrett-Crane Intertwiners}

Let me briefly discuss the formal structure of the Barrett-Crane intertwiner
of the $SO(4,C)$ general relativity for the purpose of the developing spin
foam models for real general relativity theories. It has the following elements:

\begin{itemize}
\item A gauge group $G,$

\item A homogenous space $X$ of $G,$

\item A $G$ invariant measure on $X$ and,

\item A complete orthonormal set of functions which call as $T-$functions
which are maps from $X$ to the Hilbert spaces of a subset of unitary
representations of $G$:
\[
T_{\rho}:X\rightarrow D_{\rho},
\]
where $\rho$ is a representation of $G.$ The $T-$functions correspond to the
various unitary representations under the transformation of $X$ under $G.$ The
$T-$functions are complete in the sense that on the $L^{2}$ functions on $X$
they define invertible Fourier transforms. The $T-$ functions are written
using its components in a linear vector basis of representation $D_{\rho}$.
\end{itemize}

Formally Barrett-Crane intertwiners are quantum states $\Psi$ associated to
closed simplicial two surfaces defined as an integral of a outer product of
$T-$functions on the space $X$:
\[
\Psi=\int\limits_{X}\prod\limits_{\otimes\rho}T_{\rho}(x)d_{X}x\in
\prod\limits_{\otimes\rho}D_{\rho}\text{.}%
\]
It can seen that $\Psi$ is gauge invariant under $G$ because of the invariance
of the measure $d_{X}x$.

\subsection{The Real Barrett-Crane Models}

Consider a four-simplex with complex bivectors $B_{i}$, $i=1$ to $10$
associated with its triangles. The discrete equivalent of the area metric
reality constraint is the bivector scalar product reality constraint. Then the
bivector scalar product reality constraint requires%
\[
\operatorname{Im}(B_{i}\wedge B_{j})=0~~~\forall i,j.
\]

I\ would like to formally reduce the Barrett-Crane models for real general
relativity from that of the $SO(4,C)$ Barrett-Crane model by using the
bivector scalar products reality constraint. Precisely I\ plan to use the
following three ideas to reduce the Barrett-Crane models:

\begin{enumerate}
\item The formal structure of the reduced intertwiners should be the same as
that of the $SO(4,C)$ Barrett-Crane model,

\item The eigen value of the Casimir corresponding to the square of the area
of any triangle must be real. I\ would like to refer to this as the
self-reality constraint\footnote{I\ would like to mention that the areas being
real necessarily does not mean that the bivectors must also be real.},

\item The eigen values of the square of area Casimir corresponding to the
representations associated with the internal links of the intertwiner must be
real. I\ would like to refer to this as the cross-reality constraint.
\end{enumerate}

The first idea sets a formal ansatz for the reduction process. The simple and
symmetric nature of the $SO(4,C)$ (or $SO(4,R)$) Barrett-Crane intertwiner
and, the work done in Ref:\cite{SimpleSpinNetworks}, Ref:\cite{ReisenBCinter}
and Ref:\cite{SytematicDerReim} can be considered as evidences for the formal
structure to be the general form of structure of intertwiners for all
signatures. Here we assume this idea as a hypothesis.

The square of the area of a triangle is simply the scalar product of the
bivector of a triangle with itself. Second condition is the quantum equivalent
of the reality of the scalar product of a bivector associated with a triangle
with itself. Once the second condition is imposed the third condition is the
quantum equivalent of the reality of the scalar product of the two bivectors
of any two triangle of a tetrahedron\footnote{We have ignored to impose
reality of the scalar products of the bivectors associated to any two
triangles of the same four simplex which intersect at only at one vertex. This
is because these constraints appears not to be needed for a formal extraction
of the Barrett-Crane models of real general relativity from that of $SO(4,C)$
general relativity described in this section. Imposing these constraints may
not be required because of the enormous redundancy in the bivector scalar
product reality constraints. This issue need to be carefully investigated}.

My goal is to use the above principles to derive reduced Barrett-Crane models
and later one can convince oneself by identifying and verifying that the
Barrett-Crane constraints are satisfied for a subgroup of $SO(4,C)$ for each
of the reduced model.

In general by reducing a certain Hilbert space associated with the
representations of a group $G$ by some constraints, the resultant Hilbert
space need not contain the states gauge invariant under $G$. In that case one
can look for gauge invariance states under subgroups of $G$. In our case we
will find that the suitable quantum states extracted by adhering to the above
principles are gauge symmetry reduced versions of $SO(4,C)$ Barrett-Crane
states. They are gauge invariant only under the real subgroups of $SO(4,C)$.

Let $P$ be a formal projector which reduces the Hilbert space $D_{\chi_{L}%
}\otimes$ $D_{\chi_{R}}$ to a reduced Hilbert space such that the reality
constraints are satisfied. Let me assume as an ansatz that now the complex
three sphere is replaced by its subspace $X$ due to projection. Now I expect,
the projected $SO(4,C)$ Barrett-Crane intertwiner is spanned by the following
states for all $\chi_{i}$ satisfying the reality constraints:%
\[
\Psi_{X}=\int_{x\in X}%
%TCIMACRO{\dprod \limits_{i}}%
%BeginExpansion
{\displaystyle\prod\limits_{i}}
%EndExpansion
PT_{\chi_{i}}(\mathfrak{g}(x))\tilde{d}\mathfrak{g}(x),
\]
where $\tilde{d}\mathfrak{g}(n)$ is the reduced measure of $d\mathfrak{g}(n)$
on $X.$ The imposition of the self-reality constraints expressed at the
quantum level sets $\rho_{i}$ or $n_{i}$ to be zero on each vertex of the
$SO(4,C)$ Barrett-Crane intertwiner. Let me rewrite the projected intertwiner
as follows.%
\[
\Psi_{X}=\int_{x,y\in X}%
%TCIMACRO{\dprod \limits_{1,2}}%
%BeginExpansion
{\displaystyle\prod\limits_{1,2}}
%EndExpansion
PT_{\chi_{1}}(\mathfrak{g}(x))\delta_{X}(x,y)%
%TCIMACRO{\dprod \limits_{3,4}}%
%BeginExpansion
{\displaystyle\prod\limits_{3,4}}
%EndExpansion
PT_{\chi_{1}}(\mathfrak{g}(y))d^{X}g(x)d^{X}g(y),
\]
where $\delta_{X}(x,y)$ is the delta function on $X$. Since $X$ is a subspace
of $SL(2,C)$ a harmonic expansion can be derived for $\delta(x,y)$ using the
unitary representations of $SL(2,C)$. Since the intertwiner must obey the
cross reality constraint the harmonic expansion must only contain simple
representations of $SL(2,C)$ ($\rho$ or $n$ is zero).

For the Fourier transform defined by $PT_{\chi}(\mathfrak{g}(x))$ to be
complete and orthonormal I must have%
\[
\int_{\chi\in Q}\bar{\chi}\chi tr(PT_{\chi}(\mathfrak{g}(x))PT_{\chi
}(\mathfrak{g}(y)))d\chi=\delta_{X}(x,y),
\]
where $Q$ is the set of all simple representations\footnote{One could also
call the simple representations of $SL(2,C)$ as the real representations since
it corresponds to the real areas and the real homogenous spaces. But I\ will
avoid this to avoid any possible confusion.} of $SL(2,C)$ required for the
expansion. Only the simple representations of $SL(2,C)$ must be used to
satisfy the cross-reality constraints. Thus, the number of reduced
intertwiners derivable is directly related to the possible solutions for this
equation (subjected to Barrett-Crane constraints).

The equation of a complex three sphere is
\[
x^{2}+y^{2}+z^{2}+t^{2}=1.
\]
There are four different topologically different maximally connected real
subspaces of $CS^{3}$ such that the harmonic (Fourier) expansions on these
spaces use the simple representations of $SL(2,C)$ only. They are namely, the
three sphere $S^{3},$ the real hyperboloid $H^{+}$, the imaginary hyperboloid
$H^{-}$ and the Kleinien hyperboloid\footnote{By Kleinien hyperboloid I\ refer
to the space described by $x^{2}+y^{2}-z^{2}-t^{2}=1$ for real $x,y,z$ and
$t.$} $K^{3}$. Each of these subspace $X$ are maximal real subspaces of
$CS^{3}$. They are all homogenous under the action of a maximal real
subgroup\footnote{The real group is maximal in the sense that there is no
other real topologically connected subgroup of $SO(4,C)$ that is bigger.}
$G_{X}$ of $SO(4,C)$. There exists a $G_{X}$ invariant measure $d^{X}(x)$. The
reduced bivectors acting on the functions on $X$ effectively take values in
the Lie algebra of $G_{X}$. Since the measure $d^{X}(n)$ is invariant, the
reduced intertwiner is gauge invariant. So the intertwiner $\Psi_{X}$ must
correspond to the quantum general relativity for the group $G_{X}$.

Let the coordinates of $n=(x,y,z,t)$ be restricted to real values here after
in this section. Let me discuss the various reduced intertwiners:

\begin{enumerate}
\item $\rho=0$ case: This uses only the $\chi=(0,n)$ representations only.
This corresponds to $X=S^{3}$, satisfying%
\[
x^{2}+y^{2}+z^{2}+t^{2}=1,
\]
which is invariant under $SO(4,R)$. So this case corresponds to the Riemannian
general relativity. The appropriate projected $T-$functions are the
representation matrices of $SU(2)\approx S^{3}$ and the reduced measure is the
Haar measure of $SU(2).$ The intertwiner I get is the Barrett-Crane
intertwiner for the Riemannian general relativity. Here the $\chi^{\prime}s$
has been replaced by the $J^{\prime}s$ and the complex three sphere by the
real three sphere. The case of going from the $SO(4,C)$ Barrett-Crane model to
the Riemannian Barrett-Crane model is intuitive. It is a simple process of
going from complex three sphere to its subspace the real three sphere.

\item $n=0$ case: This uses $\chi=(\rho,0)$ representations only: This
corresponds to $X$ as a space-like hyperboloid (only one sheet) with
$G_{X}=SO(3,1,R)$:%
\[
x^{2}+y^{2}+z^{2}-t^{2}=1.
\]

The intertwiner now corresponds to the Lorentzian general relativity. This
intertwiner was introduced in $\cite{BCLorentzian}$. The unitary
representations of the Lorentz group on the real hyperboloid have been studied
by Gelfand and Naimarck \cite{IMG}, from which the $T-$functions are
\[
T_{\rho}(x)[\xi]=[\xi.x]^{\frac{1}{2}i\rho-1},
\]
where $\xi$ $\in$ null cone intersecting $t=1$ plane in the Minkowski space.
Here $\xi$ replaces $(z_{1},z_{2})$ in the $T-$function $T_{\chi}%
(\mathfrak{g}(x))(z_{1},z_{2})$of the $SO(4,C)$ Barrett-Crane Model. An
element $g$ $\in$ $SO(3,1)$ acts as a shift operator as follows:
\begin{subequations}
\label{eq.action}%
\begin{align*}
gT_{\rho}(x)[\xi]  &  =T_{\rho}(gx)[\xi]\\
&  =T_{\rho}(x)[g^{-1}\xi].
\end{align*}

This intertwiner was first introduced in $\cite{BCLorentzian}$.

\item Combination of $(0,n)\ $and$(\rho,0)$ representations: There are two
possible models corresponding to this case. One of them has $X\ $as the
Kleinien hyperboloid defined by%
\end{subequations}
\[
x^{2}+y^{2}-z^{2}-t^{2}=1,
\]

with $G_{X}=SO(2,2,R)$. Here the $X$ is isomorphic to $SU(1,1)\approx
SL(2,R)$. The intertwiner now corresponds to Kleinien general relativity (
$++--$ signature). The $T-$functions are of the form $T_{\chi}(\mathrm{k}%
(n))(z_{1},z_{2})$ where $z_{1}$ and $z_{2}$ takes real values only (please
refer to appendix $C$ ), $\chi\neq0$ and $\mathrm{k}$ is an isomorphism from
the Kleinien hyperboloid to $SU(1,1)$ defined by
\[
\mathrm{k}(n)=\left[
\begin{array}
[c]{cc}%
x-iy & z-it\\
z+it & x+iy
\end{array}
\right]  .
\]
The representations corresponding to the $n=0$ and $\rho=0$ cases are
qualitatively different. The representations corresponding to $\rho\neq0$ are
called the \textbf{continuous representations} and those to $n\neq0$ are
called the \textbf{discrete representations. }The action of $g\in SO(2,2,R)$
on the $T-$functions is
\[
gT_{\chi}(\mathrm{k}(x))=T_{\chi}(\mathrm{k}(g(x)),
\]
where $g(x)$ is the result of action of $g$ on $x\in X$.

\item The second model using both $(0,n)\ $and$(\rho,0)$ representations: This
corresponds to the time-like hyperboloid with $G_{X}=$ $SO(3,1)$,%
\[
x^{2}-y^{2}-z^{2}-t^{2}=1,
\]
where two vectors that differ just by a sign are identified as a single point
of the space $X$. The corresponding spin foam model has been introduced by
Barrett and Crane \cite{BCReimmanion}. It has been derived using a field
theory over group formalism by Rovelli and Perez \cite{RoPeModel}. Similar to
the previous case, I have both continuous and discrete representations, with
the $T-$functions given by%
\begin{align*}
T_{\rho}(x)[\xi]  &  =[\xi.x]^{\frac{1}{2}i\rho-1},\\
T_{n}(x)[l(a,\xi)]  &  =\exp(-2in\theta)\delta(a.\xi),
\end{align*}
where the $l(a,\xi)$ is an isotropic line\footnote{A line on an imaginary
hyperboloid \cite{IMG} is the intersection of a 2-plane of the Minkowski space
with it. The line is called isotropic if the Lorentzian distance between any
two points on it is zero. An isotropic line $l$ is described by the equation
$x=s\xi+x_{0},$ $x$ is the variable point on $l$, $x_{0}$ is any fixed point
on $l,$ and $\xi$ is a null-vector. For more information please refer to
\cite{IMG}} on the imaginary hyperboloid along direction $\xi$ going through a
point $a$ on the hyperboloid and the $\theta$ is the distance between
$l(a,\xi)$ and $l(x,\xi)$ given by $\cos\theta=a.x,$ where the dot is the
Lorentzian scalar product. I have for $g\in SO(3,1,R),$%
\begin{align*}
gT_{n}(x)[l(a,\xi)]  &  =T_{n}(x)[l(a,g\xi)]\\
&  =T_{n}(g^{-1}x)[l(a,\xi)],
\end{align*}
and the action of $g$ on continuous representations are defined similar to
equation (\ref{eq.action}). The corresponding spin foam model has been
introduced and investigated before by Rovelli and Perez \cite{RoPeModel}.
\end{enumerate}

In the table below the representations and the homogenous spaces associated
with various Barrett-Crane intertwiners (models) in four dimensions have been summarized.

\begin{center}%
\begin{tabular}
[c]{|l|l|l|}\hline
$\text{\textbf{Model}}$ & $\text{Representations}$ & $\text{Homogenous Space}%
$\\\hline
$SO(4,C)~\text{model}$ & $\chi_{L}=\pm\chi_{R}=\frac{n}{2}+i\rho$ & Complex
three-sphere\\\hline
$SO(4,R)~\text{model}$ & $\rho=0\text{ Discrete irreps}$ & Real
three-sphere\\\hline
$SO(3,1)~\text{model}$ & $n=0\text{ Continuos irreps}$ & $\text{Space-Like
Hyperboloid}$\\\hline
$SO(3,1)~\text{model}$ & $n=0\oplus\rho=0$ & $\text{Time-Like Hyperboloid}%
$\\\hline
$SO(2,2)~\text{model}$ & $n=0\oplus\rho=0$ & $\text{Kleinien Hyperboloid}%
$\\\hline
\end{tabular}

\end{center}

From the table, it can be clearly seen that the representations used for the
intertwiners for real general relativity are various possible combinations of
representations of $SL(2,C)$ simply restricted by the reality condition
$n\rho=0$. Also all the homogenous spaces of the intertwiners of the real
general relativity theories are simply the all possible real cross-sections
(maximal) of the complex three sphere. From this point of view the
intertwiners for the real general relativity theories listed in the table
makes a complete set.

\subsection{The Area Eigenvalues}

Using the $T-$functions described above, the intertwiners for real general
relativity can be constructed. Using these intertwiners, spin foam models
(Barrett-Crane) for the real general relativity theories of the various
different signatures can be constructed. The square of the area of a triangle
of a four-simplex for all signatures associated with a representation $\chi$
is described by the same formula\footnote{Please refer to the end of appendix
$C$ regarding the differences between the Casimers of $SL(2,C)$ and
\ $SU(1,1).$},%
\begin{align*}
\eta_{IK}\eta_{JL}\hat{B}^{IJ}\hat{B}^{KL}  &  =\left(  \chi^{2}-1\right)
\hat{I}\\
&  =\left(  \frac{n^{2}}{2}-\rho^{2}-1\right)  \hat{I},
\end{align*}
where only of $n$ and $\rho$ is non-zero. The square of the area is negative
or positive depending on whether $\rho$ or $n$ is non-zero. The negative
(positive) sign corresponds to a time-like (space-like) area.

\section{Acknowledgement.}

I thank Allen Janis, George Sparling and John Baez for correspondences.

\appendix

\section{Unitary Representations of SL(2,$\boldsymbol{C}$)}

The Representation theory of $SL(2,\boldsymbol{C})$ was developed by Gelfand
and Naimarck \cite{IMG}. Representation theory of $SL(2,C)$ can be developed
using functions on $C^{2}$ which are homogenous in their
arguments\footnote{These functions need not be holomorphic but infinitely
differentiable may be except at the origin $(0,0)$.}. The space of functions
$D_{\chi}$ is defined as functions $f(z_{1},z_{2})$ on $C^{2}$ whose
homogeneity is described by%
\[
f(az_{1},az_{2})=a^{\chi_{1}-1}a^{\chi_{2}-1}f(z_{1},z_{2}),
\]
for all $a\neq0,$ where $\chi$ is a pair $(\chi_{1},\chi_{2})$. The linear
action of $SL(2,C)$ on $C^{2}$ defines a representation of $SL(2,C)$ denoted
by $T_{\chi}$. Because of the homogeneity of functions of $D_{\chi},$ the
representations $T_{\chi}$ can be defined by its action on the functions
$\phi(z)$ of one complex variable related to $f(z_{1},z_{2})\in$ $D_{\chi}$ by%
\[
\phi(z)=f(z,1).
\]
There are two qualitatively different unitary representations of $SL(2,C)$:
the principal series and the supplementary series, of which only the first one
is relevant to quantum general relativity. The principal unitary irreducible
representations of $SL(2,\boldsymbol{C})$ are the infinite dimensional. For
these $\chi_{1}=-\bar{\chi}_{2}=\frac{n+i\rho}{2},$ where $n$ is an integer
and $\rho$ is a real number. In this article I\ would like to label the
representations by a single complex number $\chi=\frac{n}{2}+i\frac{\rho}{2}$,
wherever necessary. The $T_{\chi}$ representations are equivalent to
$T_{-\chi}$ representations \cite{IMG}.

Let $g$ be an element of $SL(2,\boldsymbol{C})$ given by%
\[
g=\left[
\begin{array}
[c]{cc}%
\alpha & \beta\\
\gamma & \delta
\end{array}
\right]  ,
\]
where $\alpha$,$\beta$,$\gamma$ and $\delta$ are complex numbers such that
$\alpha\delta-\beta\delta=1$. Then the $D\chi$ representations are described
by the action of a unitary operator $T_{\chi}(g)$ on the square integrable
functions $\phi(z)$ of a complex variable $z$ as given below:%
\begin{equation}
T_{\chi}(g)\phi(z)=(\beta z_{1}+\delta)^{\chi-1}(\bar{\beta}\bar{z}_{1}%
+\bar{\delta})^{-\bar{\chi}-1}\phi(\frac{\alpha z+\gamma}{\beta z+\delta}).
\label{rep}%
\end{equation}
This action on $\phi(z)$ is unitary under the inner product defined by%
\[
\left(  \phi(z),\eta(z)\right)  =\int\bar{\phi}(z)\eta(z)d^{2}z,
\]
where $d^{2}z=\frac{i}{2}dz\wedge d\bar{z}$ and I\ would like to adopt this
convention everywhere. Completing $D_{\chi}$ with the norm defined by the
inner product makes it into a Hilbert space $H_{\chi}$.

Equation (\ref{rep}) can also be written in kernel form
\cite{RovPerGFTLorentz},%
\[
T_{\chi}(g)\phi(z_{1})=\int T_{\chi}(g)(z_{1},z_{2})\phi(z_{2})d^{2}z_{2},
\]
Here $T_{\chi}(g)(z_{1},z_{2})$ is defined as%
\begin{equation}
T_{\chi}(g)(z_{1},z_{2})=(\beta z_{1}+\delta)^{\chi-1}(\bar{\beta}\bar{z}%
_{1}+\bar{\delta})^{-\bar{\chi}-1}\delta(z_{2}-g(z_{1})), \label{eq.rep}%
\end{equation}
where $g(z_{1})=\frac{\alpha z_{1}+\gamma}{\beta z_{1}+\delta}$. The Kernel
$T_{\chi}(g)(z_{1},z_{2})$ is the analog of the matrix representation of the
finite dimensional unitary representations of compact groups. An infinitesimal
group element, $a$, of $SL(2,\boldsymbol{C})$ can be parameterized by six real
numbers $\varepsilon_{k}$ and $\eta_{k}$ as follows \cite{Ruhl}:%
\[
a\approx I+\frac{i}{2}\sum_{k=1}^{3}(\varepsilon_{k}\sigma_{k}+\eta_{k}%
i\sigma_{k}),
\]
where the $\sigma_{k}$ are the Pauli matrices. The corresponding six
generators of the $\chi$ representations are the $H_{k}$ and the $F_{k}$. The
$H_{k}$ correspond to rotations and the $F_{k}$ correspond to boosts. The
bi-invariant measure on $SL(2,C)$ is given by
\[
dg=\left(  \frac{i}{2}\right)  ^{3}\frac{d^{2}\beta d^{2}\gamma d^{2}\delta
}{\left\vert \delta\right\vert ^{2}}=\left(  \frac{i}{2}\right)  ^{3}%
\frac{d^{2}\alpha d^{2}\beta d^{2}\gamma}{\left\vert \alpha\right\vert ^{2}}.
\]
This measure is also invariant under inversion in $SL(2,\boldsymbol{C})$. The
Casimir operators for $SL(2,C$ $)$ are given by%
\[
\hat{C}=\det\left[
\begin{array}
[c]{cc}%
\hat{X}_{3} & \hat{X}_{1}-i\hat{X}_{2}\\
\hat{X}_{1}+i\hat{X}_{2} & -\hat{X}_{3}%
\end{array}
\right]
\]
and its complex conjugate $\bar{C}$ where $X_{i}=F_{i}+iH_{i}.$ The action of
$C$ ($\bar{C}$) on the elements of $D_{\chi}$ reduces to multiplication by
$\chi_{1}^{2}-1$ ($\chi_{2}^{2}-1$).The real and imaginary parts of $C$ are
another way of writing the Casimirs. On $D_{\chi}$ they reduce to the
following%
\begin{align*}
\operatorname{Re}(\hat{C})  &  =\left(  -\rho^{2}+\frac{n}{4}^{2}-1\right)
\hat{I},\\
\operatorname{Im}(\hat{C})  &  =\rho n\hat{I}.
\end{align*}

The Fourier transform theory on $SL(2,\boldsymbol{C})$ was developed in
Ref:\cite{IMG}. If $f(g)$ is a square integrable function on the group, it has
a group Fourier transform defined by%
\begin{equation}
F(\chi)=\int f(g)T_{\chi}(g)dg, \label{Four}%
\end{equation}
where is $F(\chi)$ is linear operator defined by the kernel $K_{\chi}%
(z_{1},z_{2})$ as follows:%
\[
F(\chi)\phi(z)=\int K_{\chi}(z,\acute{z})\phi(\acute{z})d^{2}\acute{z}.
\]
The associated inverse Fourier transform is%
\begin{equation}
f(g)=\frac{1}{8\pi^{4}}\int Tr(F(\chi)T_{\chi}(g^{-1}))\chi\bar{\chi}d\chi,
\label{invFour}%
\end{equation}
where the $\int d\chi$ indicates the integration over $\rho$ and the summation
over $n.$ From the expressions for the Fourier transforms, I can derive the
orthonormality property of the $T_{\chi}$ representations,%
\[
\int_{SL(2,C)}T_{\acute{z}_{1}\chi_{1}}^{z_{1}}(g)T_{\acute{z}_{2}\chi_{2}%
}^{\dag z_{2}}(g)dg=\frac{8\pi^{4}}{\chi_{1}\bar{\chi}_{1}}\delta(\chi
_{1}-\chi_{2})\delta(z_{1}-\acute{z}_{1})\delta(z_{2}-\acute{z}_{2}),
\]
where $T_{\chi}^{\dagger}$ is the Hermitian conjugate of $T_{\chi}$.

The Fourier analysis on $SL(2,C)$ can be used to study the Fourier analysis on
the complex three sphere $CS^{3}$. If $x=(a,b,c,d)\in$ $CS^{3}$ then the
isomorphism $\mathfrak{g}:CS^{3}\longrightarrow SL(2,C)\ $can be defined by
the following:%
\[
\mathfrak{g}(x)=\left[
\begin{array}
[c]{cc}%
a+ib & c+id\\
-c+id & a-ib
\end{array}
\right]  .
\]
Then, the Fourier expansion of $f(x)$ $\in L^{2}(CS^{3})$ is given by%
\[
f(x)=\frac{1}{8\pi^{4}}\int Tr(F(\chi)T_{\chi}(\text{$\mathfrak{g}$}%
(x)^{-1})\chi\bar{\chi}d\chi
\]
and its inverse is
\[
F(\chi)=\int f(g)T_{\chi}(\mathfrak{g}(x))dx,
\]
where the $dx$ is the measure on $CS^{3}$. The measure $dx$ is equal to the
bi-invariant measure on $SL(2,C)$ under the isomorphism $\mathfrak{g}$.

The expansion of the delta function on $SL(2,C)$ from equation (\ref{invFour})
is%
\begin{equation}
\delta(g)=\frac{1}{8\pi^{4}}\int tr\left[  T_{\chi}(g)\right]  \chi\bar{\chi
}d\chi. \label{deltaExp}%
\end{equation}
Let me calculate the trace $tr\left[  T_{\chi}(g)\right]  $. If $\lambda
=e^{\rho+i\theta}$ and $\frac{1}{\lambda}$ are the eigen values of $g$ then%
\[
tr\left[  T_{\chi}(g)\right]  =\dfrac{\lambda^{\chi_{1}}\bar{\lambda}%
^{\chi_{2}}+\lambda^{-\chi_{1}}\bar{\lambda}^{-\chi_{2}}}{\left\vert
\lambda-\lambda^{-1}\right\vert ^{2}},
\]
which is to be understood in the sense of distributions \cite{IMG}. The trace
can be explicitly calculated as%
\begin{equation}
tr\left[  T_{\chi}(g)\right]  =\dfrac{\cos(\eta\rho+n\theta)}{2\left\vert
\sinh(\eta+i\theta)\right\vert ^{2}}. \label{eq.trsl(2,C)}%
\end{equation}
Therefore, the expression for the delta on $SL(2,C)$ explicitly is%
\begin{equation}
\delta(g)=\frac{1}{8\pi^{4}}\sum_{n}\int d\rho(n^{2}+\rho^{2})\dfrac{\cos
(\rho\eta+n\theta)}{\left\vert \sinh(\eta+i\theta)\right\vert ^{2}}.
\label{deltaExplicit}%
\end{equation}
Let us consider the integrand in equation (\ref{invFour}). Using equation
(\ref{Four}) in it we have
\begin{align}
Tr(F(\chi)T_{\chi}(g^{-1}))\chi\bar{\chi}  &  =\chi\bar{\chi}\int f(\acute
{g})Tr(T_{\chi}(\acute{g})T_{\chi}(g^{-1}))d\acute{g}\nonumber\\
&  =\chi\bar{\chi}\int f(\acute{g})Tr(T_{\chi}(\acute{g}g^{-1}))d\acute{g}.
\label{SignChi}%
\end{align}
But, since the trace is insensitive to an overall sign of $\chi$, so are the
terms of the Fourier expansion of the $L^{2}$ functions on $SL(2,C)\ $and
$CS^{3}$.

\section{Unitary Representations of $SO(4,C)$}

The group $SO(4,C)$ is related to its universal covering group $SL(2,C)\times
SL(2,C)$ by the relationship $SO(4,C)\approx\frac{SL(2,C)\times SL(2,C)}%
{Z^{2}}$. The map from $SO(4,C)$ to $SL(2,C)\times SL(2,C)$ is given by the
isomorphism between complex four vectors and $GL(2,C)$ matrices. If
$X=(a,b,c,d)$ then $G:C^{4}\longrightarrow GL(2,C)\ $can be defined by the
following:%
\[
G(X)=\left[
\begin{array}
[c]{cc}%
a+ib & c+id\\
-c+id & a-ib
\end{array}
\right]  .
\]
It can be easily inferred that $\det G(X)=a^{2}+$ $b^{2}+c^{2}+d^{2}$ is the
Euclidean norm of the vector $X$. Then, in general a $SO(4,C)$ rotation of a
vector $X$ to another vector $Y$ is given in terms of two arbitrary
$SL(2,C)\ $matrices $g_{L~B}^{~~A},~g_{R~B^{^{\prime}}}^{~~A^{^{\prime}}}\in
SL(2,C)$ by
\[
G(Y)^{AA^{^{\prime}}}=g_{L~B}^{~~A}g_{R~B^{^{\prime}}}^{~A^{^{\prime}}}%
G^{AB}(X),
\]
where $G^{AB}(X)$ is the matrix elements of $G(X)$. The above transformation
does not differentiate between $(L_{B}^{A},R_{B^{^{\prime}}}^{A^{^{\prime}}})$
and $(-L_{B}^{A},-R_{B^{^{\prime}}}^{A^{^{\prime}}})$ which is responsible for
the factor $Z_{2}$ in $SO(4,C)\approx\frac{SL(2,C)\times SL(2,C)}{Z^{2}}$.

The unitary representation theory of the group $SL(2,C)\times SL(2,C)$ is
easily obtained by taking the tensor products of two Gelfand-Naimarck
representations of $SL(2,C)$. The Fourier expansion for any function
$f(g_{L},g_{R})$ of the universal cover is given by%
\[
f(g_{L},g_{R})=\frac{1}{64\pi^{8}}\int\chi_{L}\bar{\chi}_{L}\chi_{R}\bar{\chi
}_{R}F(\chi_{L},\chi_{R})T_{\chi}(g_{L}^{-1})T_{\chi}(g_{R}^{-1})d\chi
_{L}d\chi_{R},
\]
where $\chi_{L}=\frac{n_{L}+i\rho_{L}}{2}$ and $\chi_{R}=\frac{n_{R}+i\rho
_{R}}{2}$. The Fourier expansion on $SO(4,C)$ is given by reducing the above
expansion such that $f(g_{L},g_{R})=f(-g_{L},-g_{R})$. From equation
(\ref{eq.trsl(2,C)}) I have%
\[
tr\left[  T_{\chi}(-g)\right]  =(-1)^{n}tr\left[  T_{\chi}(-g)\right]  ,
\]
where $\chi=\frac{n+i\rho}{2}$. Therefore%
\[
f(-g_{L},-g_{R})=\frac{1}{8\pi^{4}}\int\chi_{L}\bar{\chi}_{L}\chi_{R}\bar
{\chi}_{R}F(\chi_{L},\chi_{R})(-1)^{n_{L}+n_{R}}T_{\chi}(g_{L}^{-1})T_{\chi
}(g_{R}^{-1})d\chi_{L}d\chi_{R}.
\]
This implies that for $f(g_{L},g_{R})=f(-g_{L},-g_{R}),$ I must
have$~(-1)^{n_{L}+n_{R}}$ $=1$. From this, I can infer that the representation
theory of $SO(4,C)$ is deduced from the representation theory of
$SL(2,C)\times SL(2,C)$ by restricting $n_{L}+n_{R}$ to be even integers. This
means that $n_{L}$ and $n_{R}$ should be either both odd numbers or even
numbers. I would like to denote the pair $(\chi_{L},\chi_{R})$ ($n_{L}+n_{R}$
even) by $\omega$.

There are two Casimir operators available for $SO(4,C),$ namely $\varepsilon
_{IJKL}\hat{B}^{IJ}\hat{B}^{KL}$ and $\eta_{IK}\eta_{JL}\hat{B}^{IJ}\hat
{B}^{KL}$. The elements of the representation space $D_{\chi_{L}}\otimes$
$D_{\chi_{R}}$ are the eigen states of the Casimirs. On them, the operators
reduce to the following:
\begin{equation}
\varepsilon_{IJKL}\hat{B}^{IJ}\hat{B}^{KL}=\frac{\chi_{L}^{2}-\chi_{R}^{2}}%
{2}~~\text{and}%
\end{equation}%
\begin{equation}
\eta_{IK}\eta_{JL}\hat{B}^{IJ}\hat{B}^{KL}=\frac{\chi_{L}^{2}+\chi_{R}^{2}%
-2}{2}.
\end{equation}

\section{Unitary Representations of $SU(1,1)$}

The unitary representations of $SU(1,1)\approx SL(2,R)$, given in
Ref:\cite{NJaVAUK}, is defined similar to that of $SL(2,C)$. The main
difference is that the $D_{\chi}$ are now functions $\phi(z)$ on $C^{1}$. The
representations are indicated by a pair $\chi=$ $(\tau,\varepsilon),$
$\varepsilon$ is the parity of the functions ($\varepsilon$ is $0$ for even
functions and $\frac{1}{2}$ for odd functions) and $\tau$ is a complex number
defining the homogeneity:%
\[
\phi(az)=\left\vert a\right\vert ^{2\tau}sgn(a)^{2\varepsilon}\phi(z),
\]
where $a$ is a real number. Because of homogeneity the $D_{\chi}$ functions
can be related to the infinitely differentiable functions $\phi(e^{i\theta})$
on $S^{1}$ where $\theta$ is the coordinate on $S^{1}$. The representations
are defined by%
\begin{equation}
T_{\chi}(g)\phi(e^{i\theta})=(\beta e^{i\theta}+\bar{\alpha})^{\tau
+\varepsilon}(\bar{\beta}e^{-i\theta}+\alpha)^{\tau-\varepsilon}\phi
(\frac{\alpha z+\bar{\beta}}{\beta z+\bar{\alpha}}). \label{T-SL(2,R)}%
\end{equation}
There are two types of the unitary representations that are relevant for
quantum general relativity: the continuous series and the discrete series. For
the continuous series $\chi=(i\rho-\frac{1}{2},\varepsilon),$ where $\rho$ is
a non-zero real number. Let me denote the continuous series representations
with suffix or prefix $c,$ for example $T_{\chi}^{c}$.

There are two types of discrete series representations which are indicated by
signs $\pm.$ They have their respective homogeneity as $\chi_{\pm
}=(l,\varepsilon_{l}^{\pm})$ where $\varepsilon_{l}^{\pm}=\pm1$ is defined by
the condition $l\pm\varepsilon_{l}^{\pm}$ is an integer. Let me denote the
representations as $T_{l}^{+}$ and $T_{l}^{-}$. The $T_{l}^{+}$ ($T_{l}^{-}$)
representations can be re-expressed as linear operators on the functions
$\phi_{+}(z)$($\phi_{-}(z)$) on $C^{1}$ that are analytical inside (outside)
the unit circle. The $T_{l}^{\pm}(g)$ are defined as%
\[
T_{l}^{\pm}(g)\phi_{\pm}(z)=\left\vert \beta z+\bar{\alpha}\right\vert
^{2l}\phi_{\pm}(\frac{\alpha z+\bar{\beta}}{\beta z+\bar{\alpha}}).
\]
The inner products are defined by%
\begin{align*}
(f_{1},f_{2})_{c}  &  =\frac{1}{2\pi}\int_{0}^{2\pi}f_{1}(e^{i\theta
})\overline{f_{2}(e^{i\theta})}d\theta,\\
(f_{1},f_{2})_{+}^{l}  &  =\frac{1}{\Gamma(-2l-1)}\iint\limits_{\left\vert
z\right\vert <1}(1-\left\vert z\right\vert )^{-2l-2}f_{1}(z)f_{2}%
(z)\frac{dzd\bar{z}}{2\pi i},\\
(f_{1},f_{2})_{-}^{l}  &  =\frac{1}{\Gamma(-2l-1)}\iint\limits_{\left\vert
z\right\vert >1}(1-\left\vert z\right\vert )^{-2l-2}f_{1}(z)f_{2}%
(z)\frac{dzd\bar{z}}{2\pi i}.
\end{align*}
The Fourier transforms are defined for the unitary representations by%
\begin{align*}
F_{c}(\chi)  &  =\int f(g)T_{\chi}^{c}(g)dg,\\
F_{+}(l)  &  =\int f(g)T_{l}^{+}(g)dg,~~\text{and}\\
F_{-}(l)  &  =\int f(g)T_{l}^{-}(g)dg,
\end{align*}
where $dg$ is the bi-invariant measure on the group.

The inverse Fourier transform is defined by
\[
f(g)=\frac{1}{4\pi^{2}}\left\{
\begin{array}
[c]{c}%
\sum_{l\in\frac{1}{2}N_{0}}(l+\frac{1}{2})Tr[F_{+}^{\dagger}(l)(T_{l}%
^{+}(g))+F_{-}^{\dagger}(l)(T_{l}^{-}(g))]\\
+\sum_{\varepsilon}\int_{0}^{\infty}\rho Tr[F(\chi)T_{\rho}^{\dagger}]\tanh
\pi(\rho+i\varepsilon)d\rho
\end{array}
\right\}  .
\]
The $T_{(\tau,\varepsilon)}$ is equivalent to $T_{(-\tau-1,\varepsilon)}$. The
Casimir operator for the $T_{\chi}$ representations (all) can be defined
similar to $SU(2)$ and its eigen values are%
\[
C=\tau(\tau+1),
\]
where the $\tau$ comes from $\chi=(\tau,\varepsilon).$ The $\tau$ in this
section is related to the $\chi$ in the representations of $SL(2,C)\ $by
$\chi=\tau+\frac{1}{2}$. The expressions for the Casimirs of the two groups
differ by a factor of $4$.

\section{Reality Constraint for Arbitrary Metrics}

Here we\ analyze the area metric reality constraint for a metric $g_{ac}$ of
arbitrary rank in arbitrary dimensions, with the area metric defined as
$A_{abcd}=$ $g_{a[c}g_{d]b}$. Let the rank of $g_{ac}$ be $r$.

If the rank $r=1$ then $g_{ab}$ is of form $\lambda_{a}\lambda_{b}$ for some
complex non zero co-vector $\lambda_{a}$. This implies that the area metric is
zero and therefore not an interesting case.

Let me prove the following theorem.

\begin{theorem}
If the rank $r$ of $g_{ac}$ is $\geq2,$then the area metric reality constraint
implies the metric is real or imaginary. If the rank $r$ of $g_{ac}$ is equal
to $1,$ then the area metric reality constraint implies $g_{ac}=\eta\alpha
_{a}\alpha_{b}$ for some complex $\eta\neq0$ and real non-zero co-vector
$\alpha_{a}$.
\end{theorem}

The area metric reality constraint implies%
\begin{equation}
g_{ac}^{R}g_{db}^{I}=g_{ad}^{R}g_{cb}^{I}. \label{eq.reality}%
\end{equation}
Let $g_{AC}$ be a $r$ by $r$ submatrix of $g_{ac}$ with a non zero
determinant, where the capitalised indices are restricted to vary over the
elements of $g_{AC}$ only. Now we have%
\begin{equation}
g_{AC}^{R}g_{DB}^{I}=g_{AD}^{R}g_{CB}^{I}. \label{eq.reality1}%
\end{equation}
From the definition of the determinant and the above equation we have
\[
\det(g_{AC})=\det(g_{AC}^{R})+\det(g_{AC}^{I}).
\]
Since $\det(g_{AC})\neq0$ we have either $\det(g_{AC}^{R})$ or $\det
(g_{AC}^{I})$ not equal to zero. Let me assume $g_{ac}^{R}\neq0$. Then
contracting both the sides of equation (\ref{eq.reality1}) with the inverse of
$g_{AC}^{R}$ we find $g_{DB}^{I}$ is zero. Now from equation (\ref{eq.reality}%
) we have%
\begin{equation}
g_{AC}^{R}g_{dB}^{I}=g_{Ad}^{R}g_{CB}^{I}=0. \label{eq.1}%
\end{equation}
Since the Rank of $g_{AC}^{R}\geq2$ we can always find a $g_{AC}^{R}$ $\neq0$
for some fixed $A$ and $C.$ Using this in equation (\ref{eq.1}) we find
$g_{dB}^{I}$ is zero. Now consider the following:%
\begin{equation}
g_{AC}^{R}g_{db}^{I}=g_{Ad}^{R}g_{Cb}^{I}. \label{eq.2}%
\end{equation}
we can always find a $g_{AC}^{R}\neq0$ for some fixed $A$ and $C.$ Using this
in equation (\ref{eq.2}) we find $g_{db}^{I}=0$. So we have shown that if
$g_{ac}^{R}\neq0$ then $g_{db}^{I}=0$. Similarly if we can show that if
$g_{ac}^{I}\neq0$ then $g_{db}^{R}=0$.

\section{Field Theory over Group and Homogenous Spaces.}

One of the problems with the Barrett-Crane model for general relativity is its
dependence on the discretization of the manifold. A discretization independent
model can be defined by summing over all possible discretizations. With a
proper choice of amplitudes for the lower dimensional simplices the BF\ spin
foams can be reformulated as a field theory over a group (GFT)
\cite{ooguriBFderv}. Similarly, the Barrett-Crane models can be reformulated
as a field theory over the homogenous space of the group \cite{GFTreimm}.
Consider a tetrahedron. Let a group element $g_{i}$ be associated with each
triangle $i$ of the tetrahedron. Let a real field $\phi(g_{1},g_{2},$
$g_{3},g_{4})$ invariant under the exchange of its arguments be associated
with the tetrahedron. Let the field be invariant under the simultaneous (left
or right) action of a group element $g$ on its variables. Then the kinetic
term is defined as%

\[
K.E=\int\prod\limits_{i=1}^{4}dg_{i}\phi^{2}.
\]

To define the potential term, consider a four-simplex. Let $g_{i},$ where
$i=1$ to $10$ be the group elements associated with its ten triangles. With
each tetrahedron $e$ of the four-simplex, associate a $\phi$ field which is a
function of the group elements associated with its triangles. Denote it as
$\phi_{e}$. Then the potential term is defined as%
\begin{align*}
P.E  &  =\\
&  =\frac{\lambda}{5!}\int\prod\limits_{i=1}^{10}dg_{i}\prod\limits_{e=1}%
^{5}\phi_{e},
\end{align*}
where $\lambda$ is an arbitrary constant.

Now the action for a GFT can be defined as%
\[
S(\phi)=K.E+P.E=\int\prod\limits_{i=1}^{4}dg_{i}\phi^{2}+\frac{\lambda}%
{5!}\int\prod\limits_{i=1}^{10}dg_{i}\prod\limits_{e=1}^{5}\phi_{e}.
\]
The action has two terms, namely the kinetic term and the potential terms. The
Partition function of the $GFT$ is%

\[
Z=\int D\phi e^{-S(\phi)}.
\]
Now, an analysis of this partition function yields the sum over spin foam
partitions of the four dimensional BF theory for group $G$ for all possible
triangulations. From the analysis of the GFT we can easily show that this
result is valid for $G=$ $SO(4,C)$ with the unitary representations defined in
the appendix B.

Let us assume $\phi$ is invariant only under the simultaneous action of an
element of a subgroup $H$ of $G$. Then, if $G=SO(4,R)$ and $H=SU(2)$ we get
GFTs for the Barrett-Crane model\footnote{Depending on whether we are using
the left or right action of $G\ $on $\phi$, we get two different models that
differ by amplitudes for the lower dimensional simplices \cite{GFTreimm}.}
\cite{GFTreimm}. Similarly, if $G=$ $SL(2,C)$ and $H=SU(2)$ or $SU(1,1)$, we
can define GFT for the Lorentzian general relativity \cite{RoPeModel},
\cite{RovPerGFTLorentz}. The representation theories of $SO(4,C)$ and
$SL(2,C)$ has similar structure to those of $SO(4,R)$ and $SU(2)$
respectively. So the GFT\ with $G=SO(4,C)$ and $H=SL(2,C)$ should yield the
sum over triangulation formulation of the $SO(4,C)\ $Barrett-Crane model. The
details of this analysis and its variations will be presented
elsewhere.

\end{document}